\documentclass[a4paper,UKenglish]{lipics-v2018}

\usepackage[utf8]{inputenc}
\usepackage{mathscinet}
\usepackage{comment}
\usepackage{graphicx}
\usepackage{booktabs}
\usepackage{hyperref}
\usepackage{pgfkeys}
\usepackage{enumerate}
\usepackage{verbatim}
\usepackage{nicefrac}
\usepackage{framed}

\newcommand{\lv}[1]{}

\usepackage[dvipsnames]{xcolor}
\usepackage{tikz}
\usetikzlibrary{fit}
\usetikzlibrary{intersections}
\usetikzlibrary{decorations.pathreplacing}
\usetikzlibrary{calc}

\usepackage{amsmath}
\usepackage{amssymb}
\usepackage{thmtools}
\usepackage{thm-restate}
\usepackage{xspace}

\newtheorem{question}{Open question}

\newcommand{\tg}{\textsc{Terrain Guarding}\xspace}
\newcommand{\otg}{\textsc{Orthogonal Terrain Guarding}\xspace}
\newcommand{\btg}{\textsc{(Orthogonal) Terrain Guarding}\xspace}

\tikzstyle{vp}=[circle,fill,inner sep=0pt, minimum size=0.1cm]
\tikzstyle{vps}=[circle,fill,inner sep=0pt, minimum size=0.065cm]

\title{Orthogonal Terrain Guarding is NP-complete}
\titlerunning{Orthogonal Terrain Guarding is NP-complete}

\author{\'Edouard Bonnet}{ENS Lyon, LIP\\ {Lyon, France}}{edouard.bonnet@dauphine.fr}{}{}
\author{Panos Giannopoulos}{Department of Computer Science, Middlesex University\\{London, UK}}{p.giannopoulos@mdx.ac.uk}{}{}

\authorrunning{\'E. Bonnet and P. Giannopoulos}
\Copyright{\'Edouard Bonnet and Panos Giannopoulos}
\subjclass{}

\funding{Supported by EPSRC grant FptGeom (EP/N029143/1)}

\nolinenumbers

\begin{document}

\maketitle

\begin{abstract}
  A terrain is an x-monotone polygonal curve, i.e., successive vertices have increasing x-coordinates.
  \textsc{Terrain Guarding} can be seen as a special case of the famous art gallery problem where one has to place at most $k$ guards on a terrain made of $n$ vertices in order to fully see it.
  In 2010, King and Krohn showed that Terrain Guarding is NP-complete [SODA '10, SIAM J. Comput. '11] thereby solving a long-standing open question.
  They observe that their proof does not settle the complexity of \textsc{Orthogonal Terrain Guarding} where the terrain only consists of horizontal or vertical segments; those terrains are called rectilinear or orthogonal.
  Recently, Ashok et al. [SoCG'17] presented an FPT algorithm running in time $k^{O(k)}n^{O(1)}$ for \textsc{Dominating Set} in the visibility graphs of rectilinear terrains without 180-degree vertices.
  They ask if \textsc{Orthogonal Terrain Guarding} is in P or NP-hard.
  In the same paper, they give a subexponential-time algorithm running in $n^{O(\sqrt n)}$ (actually even $n^{O(\sqrt k)}$) for the general \tg and notice that the hardness proof of King and Krohn only disproves a running time $2^{o(n^{1/4})}$ under the ETH.
  Hence, there is a significant gap between their $2^{O(n^{1/2} \log n)}$-algorithm and the no $2^{o(n^{1/4})}$ ETH-hardness implied by King and Krohn's result.

  In this paper, we adapt the gadgets of King and Krohn to rectilinear terrains in order to prove that even \textsc{Orthogonal Terrain Guarding} is NP-complete.
  Then, we show how to obtain an improved ETH lower bound of $2^{\Omega(n^{1/3})}$ by refining the quadratic reduction from \textsc{Planar 3-SAT} into a cubic reduction from \textsc{3-SAT}.
  This works for both \textsc{Orthogonal Terrain Guarding} and \textsc{Terrain Guarding}.
\end{abstract}

\section{Introduction}

\tg is a natural restriction of the well-known art gallery problem.
It asks, given an integer $k$, and an \emph{$x$-monotone polygonal chain} or \emph{terrain}, to guard it by placing at most $k$ guards at vertices of the terrain.
An \emph{$x$-monotone polygonal chain} is defined from a sequence of $n$ points of the real plane $\mathbb R^2$ $p_1=(x_1,y_1), p_2=(x_2,y_2), \ldots, p_n=(x_n,y_n)$ such that $x_1 \leqslant x_2 \leqslant \ldots \leqslant x_n$ as the succession of straight-line edges $p_1p_2, p_2p_3, \ldots, p_{n-1}p_n$.
Each point $p_i$ is called a \emph{vertex} of the terrain.
We can make each coordinate of the vertices rational without changing the (non-)existence of a solution.
We will therefore assume that the input is given as a list of $n$ pairs of rational numbers, together with the integer $k$.
A point $p$ lying on the terrain is \emph{guarded} (or seen) by a subset $S$ of guards if there is at least one guard $g \in S$ such that the straight-line segment $pg$ is entirely above the polygonal chain.
The terrain is said \emph{guarded} if every point lying on the terrain is guarded.
The visibility graph of a terrain has as vertices the geometric vertices of the polygonal chain and as edges every pair which sees each other.
Again two vertices (or points) see each other if the straight-line segment that they define is above the terrain.

The \otg is the same problem restricted to \emph{rectilinear} (also called \emph{orthogonal}) terrains, that is every edge of the terrain is either horizontal or vertical.
In other words, $p_i$ and $p_{i+1}$ share the same $x$-coordinate or the same $y$-coordinate.
We say that a rectilinear terrain is \emph{strictly rectilinear} (or \emph{strictly orthogonal}) if the horizontal and vertical edges alternate, that is, there are no two consecutive horizontal (resp.~vertical) edges.
Both problems come with two other variants: the \emph{continuous variant}, where the guards can be placed anywhere on the edges of the terrain (and not only at the vertices), and the \emph{graphic variant}, which consists of \textsc{Dominating Set} in the visibility graphs of (strictly rectilinear) terrains.
The original problem is sometimes called the \emph{discrete variant}.

It is a folklore observation that for rectilinear terrains, the discrete and continuous variants coincide.
Indeed, it is an easy exercise to show that from any feasible solution using guards in the interior of edges, one can move those guards to vertices and obtain a feasible solution of equal cardinality.
The only rule to respect is that if an edge, whose interior contained a guard, links a reflex and a convex vertex, then the guard should be moved to the reflex vertex.
We will therefore only consider \otg and \textsc{Dominating Set} in the visibility graphs of strictly rectilinear terrains.
By subdividing the edges of a strictly rectilinear terrain with an at most quadratic number of 180-degree vertices (i.e., vertices incident to two horizontal edges or to two vertical edges), one can make \emph{guarding all the vertices} equivalent to \emph{guarding the whole terrain}.
Therefore, \otg is not very different from \textsc{Dominating Set} in the visibility graph of (non necessarily strictly) rectilinear terrains (and \tg is not very different from \textsc{Dominating Set} in the visibility graph of terrains).

\subparagraph*{Exponential Time Hypothesis.}
 The \emph{Exponential Time Hypothesis} (usually referred to as the ETH) is a stronger assumption than P$\neq$NP formulated by Impagliazzo and Paturi \cite{ImpagliazzoPaturi}.
 A practical (and slightly weaker) statement of ETH is that \textsc{3-SAT} with $n$ variables cannot be solved in subexponential-time $2^{o(n)}$.
Although this is not the original statement of the hypothesis, this version is most commonly used; see also Impagliazzo et al. \cite{Impagliazzo01}.
The so-called \emph{sparsification lemma} even brings the number of clauses in the exponent.

\begin{theorem}[Impagliazzo and Paturi \cite{ImpagliazzoPaturi}]
Under the ETH, there is no algorithm solving every instance of \textsc{3-SAT} with $n$ variables and $m$ clauses in time $2^{o(n+m)}$.
\end{theorem}

As a direct consequence, unless the ETH fails, even instances with a linear number of clauses $m=\Theta(n)$ cannot be solved in $2^{o(n)}$.
Unlike P$\neq$NP, the ETH allows to rule out specific running times.
We refer the reader to the survey by Lokshtanov et al. for more information about ETH and conditional lower bounds \cite{LokshtanovMS11}.

\begin{figure}[h!]
\centering
\begin{tikzpicture}[scale=0.8]
\def\n{8}
\pgfmathsetmacro{\pn}{\n-1}
\def\s{1.5}

\foreach \i in {1,...,\n}{
    \node[draw,circle] (x\i) at (\s * \i,0) {$x_\i$} ;
}
\foreach \i [count = \j from 2] in {1,...,\pn}{
         \draw (x\i) -- (x\j) ;
}

\foreach \i/\j/\k/\l/\c in {2/3/4/1/red,1/2/4/1.5/red,4/5/6/1.5/red,6/7/8/1.5/red,4/6/8/2/red,1/4/8/2.5/red, 5/6/7/-1/blue,3/4/7/-1.5/blue,2/3/7/-2/blue, 1/2/8/-2.5/blue}{
 \draw[thick,rounded corners,\c] (x\i) -- (\s * \i, \l)  -- (\s * \k, \l) -- (x\k) ;
 \draw[thick,\c] (x\j) -- (\s * \j, \l) ;
}

\node at (0.5,1) {\textcolor{red}{$\mathcal C^+$}} ;
\node at (0.5,-1) {\textcolor{blue}{$\mathcal C^-$}} ;

\end{tikzpicture}
\caption{The bipartition $(\mathcal C^+,\mathcal C^-)$ of a \textsc{Planar 3-SAT}-instance. The three-legged arches represent the clauses.
Here is a removal ordering for $\mathcal C^-$: remove the clause on $x_5, x_6, x_7$ and its middle variable $x_6$, remove the variable $x_5$, remove the clause on $x_3, x_4, x_7$ and its middle variable $x_4$, remove the clause on $x_2, x_3, x_7$ and its middle variable $x_3$, remove the variable $x_7$, remove the clause $x_1, x_2, x_8$ and its middle variable $x_2$.}
\label{fig:planar3sat}
\end{figure}
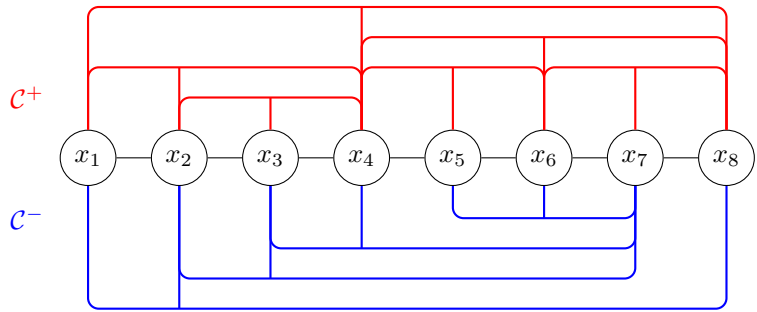

\subparagraph*{Planar satisfiability.}
\textsc{Planar 3-SAT} was introduced by Lichtenstein \cite{Lichtenstein82} who showed its NP-hardness.
It is a special case of \textsc{3-SAT} where the variable/clause incidence graph is planar even if one adds edges between two consecutive variables for a specified ordering of the variables: $x_1, x_2, \ldots, x_n$; i.e., $x_ix_{i+1}$ is an edge (with index $i+1$ taken modulo $n$).
This extra structure makes this problem particularly suitable to reduce to planar or geometric problems.
As a counterpart, the ETH lower bound that one gets from a linear reduction from \textsc{Planar 3-SAT} is worse than the one with a linear reduction from \textsc{3-SAT}; it only rules out a running time $2^{o(\sqrt n)}$.
Indeed, \textsc{Planar 3-SAT} can be solved in time $2^{O(\sqrt n)}$ and, unless the ETH fails, cannot be solved in time $2^{o(\sqrt n)}$.
A useful property of any \textsc{Planar 3-SAT}-instance is that its set of clauses $\mathcal C$ can be partitioned into $\mathcal C^+$ and $\mathcal C^-$ such that both $\mathcal C^+$ and $\mathcal C^-$ admit a \emph{removal ordering}.
A \emph{removal ordering} is a sequence of the two following deletions:
\begin{itemize}
\item(a) removing a variable which is not present in any remaining clause
\item(b) removing a clause on \emph{three consecutive remaining variables} together with the \emph{middle variable}
\end{itemize}
which ends up with an empty set of clauses.
By \emph{three consecutive remaining variables}, we mean three variables $x_i$, $x_j$, $x_k$, with $i<j<k$ such that $x_{i+1}, x_{i+2}, \ldots, x_{j-1}$ and $x_{j+1}, x_{j+2}, \ldots, x_{k-1}$ have all been removed already.
The middle variable of the clause is $x_j$.
For an example, see Figure~\ref{fig:planar3sat}.

\subparagraph*{Order claim.}
The following visibility property in a terrain made King and Krohn realize that they will crucially need the extra structure given by the planarity of \textsc{3-SAT}-instances.
\begin{lemma}[Order Claim, see Figure~\ref{fig:orderClaim}]\label{lem:orderClaim}
If $a$, $b$, $c$, $d$ happen in this order from the left endpoint of the terrain to its right endpoint, $a$ and $c$ see each other, and $b$ and $d$ see each other, then $a$ and $d$ also see each other.
\end{lemma}
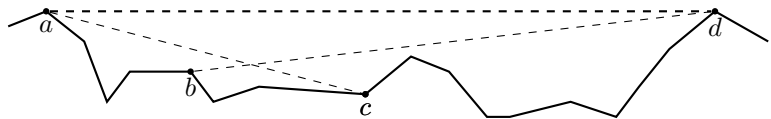
\begin{figure}[h!]
\centering
\begin{tikzpicture}
\draw[thick] (-0.3,0.8) -- (0.2,1) -- (0.7,0.6) -- (1,-0.2) -- (1.3,0.2) -- (2.1,0.2) -- (2.4,-0.2) -- (3,0) -- (4.4,-0.1) -- (5,0.4) -- (5.5,0.2) -- (6,-0.4) -- (6.3,-0.4) -- (7.1,-0.2) -- (7.7,-0.4) -- (8,0) -- (8.4,0.5) -- (9,1) -- (9.7,0.6);

\node[fill,circle,inner sep=-0.03cm] (a) at (0.2,1) {} ;
\node at (0.2,0.8) {$a$} ;

\node[fill,circle,inner sep=-0.03cm] (b) at (2.1,0.2) {} ;
\node at (2.1,0) {$b$} ;

\node[fill,circle,inner sep=-0.03cm] (c) at (4.4,-0.1) {} ;
\node at (4.4,-0.3) {$c$} ;

\node[fill,circle,inner sep=-0.03cm] (c) at (4.4,-0.1) {} ;
\node at (4.4,-0.3) {$c$} ;

\node[fill,circle,inner sep=-0.03cm] (d) at (9,1) {} ;
\node at (9,0.8) {$d$} ;

\draw[thin,dashed] (a) -- (c) ;
\draw[thin,dashed] (b) -- (d) ;
\draw[dashed,thick] (a) -- (d) ;

\end{tikzpicture}
\caption{The order claim.}
\label{fig:orderClaim}
\end{figure}
In particular, this suggests that checking in the terrain \emph{if a clause is satisfied} can only work if the encodings of the three variables contained in the clause are \emph{consecutive}.

\subparagraph*{Related work and remaining open questions for terrain guarding.}
\tg was shown NP-hard \cite{KingK11} and can be solved in time $n^{O(\sqrt k)}$ \cite{Ashok17}.
This contrasts with the parameterized complexity of the more general art gallery problem where an algorithm running in time $f(k)n^{o(k / \log k)}$ for any computable function $f$ would disprove the ETH, both for the variant \textsc{Point Guard Art Gallery} where the $k$ guards can be placed anywhere inside the gallery (polygon with $n$ vertices) and for the variant \textsc{Vertex Guard Art Gallery} where the $k$ guards can only be placed at the vertices of the polygon \cite{Bonnet16}, even when the gallery is a simple polygon (i.e., does not have holes).
\textsc{Dominating Set} on the visibility graph of strictly rectilinear terrains can be solved in time $k^{O(k)}n^{O(1)}$ \cite{Ashok17}, while it is still not known if \btg can be solved in FPT time $f(k)n^{O(1)}$ with respect to the number of guards.

There has been a succession of approximation algorithms with better and better constant ratios \cite{King06,ClarksonV07,Ben-Moshe07,Elbassioni11}.
Eventually, a PTAS was found for \tg (hence for \otg) \cite{KrohnGKV14} using local search and an idea developed by Chan and Har-Peled \cite{ChanH12} and Mustafa and Ray \cite{Mustafa10} which consists of applying the planar separator theorem to a (planar) graph relating local and global optima.
Interestingly, this planar graph is the starting point of the subexponential algorithm of Ashok et al. \cite{Ashok17}.

Again the situation is not nearly as good for the art gallery problem.
If holes are allowed in the polygon, the main variants of the art gallery problem are as hard as the \textsc{Set Cover} problem; hence a $o(\log n)$-approximation cannot exist unless P$=$NP \cite{Eidenbenz01}.
Eidenbenz also showed that a PTAS is unlikely in simple polygons \cite{Eidenbenz98}.
For simple polygons, there is a $O(\log \log OPT)$-approximation \cite{Kirkpatrick11,Kirkpatrick15} for \textsc{Vertex Guard Art Gallery}, using the framework of Br\"{o}nnimann and Goodrich to transform an $\varepsilon$-net finder into an approximation algorithm, and for \textsc{Point Guard Art Gallery} there is a randomized $O(\log OPT)$-approximation under some mild assumptions \cite{Bonnet17}, building up on \cite{Efrat06,Deshpande07}.
If a small fraction of the polygon can be left unguarded there is again a $O(\log OPT)$-approximation \cite{Elbassioni17}.
A constant-factor approximation is known for \emph{monotone polygons} \cite{KrohnN13}, where a monotone polygon is made of two terrains sharing the same left and right endpoints and except those two points the two terrains are never touching nor crossing.

The classical complexity of \otg remains the most pressing open question \cite{Ashok17}.

\begin{question}\label{question1}
Is \otg in P or NP-hard?
\end{question}

In the conclusion of the paper by Ashok et al. \cite{Ashok17}, the authors observe that the construction of King and Krohn \cite{KingK11} rules out for \tg a running time of $2^{o(n^{1/4})}$, under the ETH.
Indeed the reduction from \textsc{Planar 3-SAT} (which is not solvable in time $2^{o(\sqrt n)}$ unless the ETH fails) and its adaptation for \otg in the current paper have a quadratic blow-up: the terrain is made of $\Theta(m)=\Theta(n)$ chunks containing each $O(n)$ vertices.
On the positive side, the subexponential algorithm of Ashok et al. runs in time $2^{O(\sqrt n \log n)}$ \cite{Ashok17}.
Therefore, there is a significant gap between the algorithmic upper and lower bounds.

\begin{question}\label{question2}
Assuming the ETH, what is the provably best asymptotic running time for \tg and \otg?
\end{question}


\subparagraph*{Organization.}
In Section~\ref{sec:otg}, we address Open question~\ref{question1} by showing that \otg is also NP-hard.
We design a rectilinear subterrain with a constant number of vertices which simulates a triangular pocket surrounded by two horizontal segments.
This enables us to adapt the reduction of King and Krohn \cite{KingK11} to rectilinear terrains.
Our orthogonal gadgets make an extensive use of the \emph{triangular pockets}.

In Section~\ref{sec:eth}, we show how to make cubic reductions from \textsc{3-SAT} by refining the quadratic reductions from \textsc{Planar 3-SAT}.
This gives an improved ETH-based lower bound of $2^{\Omega(n^{1/3})}$ but does not quite resolve\footnote{In the conference version of the paper, we erroneously claim a $2^{\Omega(n^{1/2})}$ lower bound.} Open question~\ref{question2}.

\section{Orthogonal Terrain Guarding is NP-complete}\label{sec:otg}
King and Krohn give a reduction with a quadratic blow-up from \textsc{Planar 3-SAT} to \tg \cite{KingK11}.
They argue that the order claim entails some critical obstacle against straightforward hardness attempts.
In some sense, the subexponential algorithm running in time $n^{O(\sqrt n)}$ of Ashok et al. \cite{Ashok17} proves them right: unless the ETH fails, there cannot be a linear reduction from \textsc{3-SAT} to \tg.
It also justifies their idea of starting from the planar variant of \textsc{3-SAT}.
Indeed, this problem can be solved in time $2^{O(\sqrt n)}$.  

\begin{figure}[h!]
  \centering
  \begin{tikzpicture}[
      extended line/.style={shorten >=-#1},
      extended line/.default=1cm]
    \draw[rounded corners, thick] (0,0) -- (0.5,-0.2) -- (1,-0.6) -- (1.5,-1.4) -- (2,-2.3) -- (2.5,-3.5) -- (3,-5) ;
    \draw[rounded corners, thick] (7,0) -- (6.5,-0.2) -- (6,-0.6) -- (5.5,-1.4) -- (5,-2.3) -- (4.5,-3.5) -- (4,-5) ;

    \draw[red,dashed, very thin, extended line=5cm] (0,0) -- (0.5,-0.2) ;
    \draw[red,dashed, very thin, extended line=4.5cm] (6.5,-0.2) -- (6,-0.6) ;
    \draw[red,dashed, very thin, extended line=3.2cm] (1.5,-1.4) -- (2,-2.3) ;
    \draw[red,dashed, very thin, extended line=1.72cm] (5,-2.3) -- (4.5,-3.5) ;

    \node at (0.25,-0.4) {$T_2$} ;
    \begin{scope}[yshift=0.05cm,xshift=0.01cm]
      \draw[very thick] (0,0) -- (0.5,-0.2) ;
    \end{scope}

    \draw[fill,opacity=0.1] (0.01,0.05) -- (0.51,-0.15) -- (5.98,-0.55) -- (6.48,-0.15)  -- cycle;

    \node at (6.4,-0.6) {$T_1$} ;
    \begin{scope}[yshift=0.05cm,xshift=-0.02cm]
      \draw[very thick] (6.5,-0.2) -- (6,-0.6) ;
    \end{scope}

     \draw[fill,opacity=0.1] (5.98,-0.55) -- (6.48,-0.15) -- (1.53,-1.35) -- (2.03,-2.25) -- cycle;

    \node at (1.6,-2) {$T_0$} ;
    \begin{scope}[yshift=0.05cm,xshift=0.03cm]
      \draw[very thick] (1.5,-1.4) -- (2,-2.3) ;
    \end{scope}

    \draw[fill,opacity=0.1] (1.53,-1.35) -- (2.03,-2.25) -- (4.535,-3.45) -- (5.035,-2.25) -- cycle;

    \node at (5.15,-3.3) {$T_{-1}$} ;
    \begin{scope}[yshift=0.05cm,xshift=-0.035cm]
      \draw[very thick] (5,-2.3) -- (4.5,-3.5) ;
    \end{scope}

    \draw[fill,opacity=0.1] (4.535,-3.45) -- (5.035,-2.25) -- (2.54,-3.45) -- (3.04,-4.95) -- cycle;

    \node at (2.5,-4.5) {$T_{-2}$} ;
    \begin{scope}[yshift=0.05cm,xshift=0.04cm]
      \draw[very thick] (2.5,-3.5) -- (3,-5) ;
    \end{scope}
  \end{tikzpicture}
  \caption{The $V$-shaped terrain and its ordered chunks. The chunk $T_i$ only sees parts of chunks $T_{i-1}$ and $T_{i+1}$. The \emph{initial} chunk $T_0$ contains an encoding of each variable. Below this level (chunks with a negative index), we will check the clauses of $\mathcal C^-$. Above this level (chunks with a positive index), we will deal with the clauses of $\mathcal C^+$.}
  \label{fig:top-level}
\end{figure}
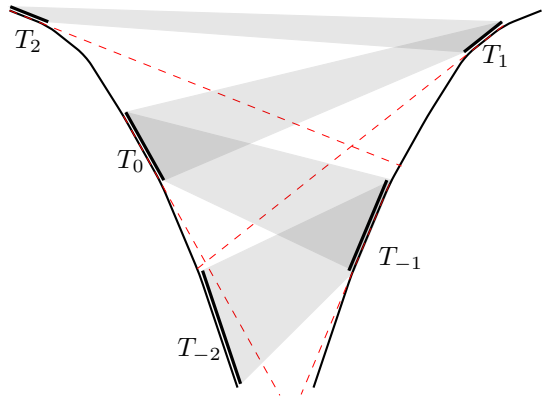

From far, King and Krohn's construction looks like a $V$-shaped terrain.
If one zooms in, one perceives that the $V$ is made of $\Theta(n)$ connected subterrains called \emph{chunks}.
If one zooms a bit more, one sees that the chunks are made of up to $n$ variable encodings each.
Let us order the chunks from bottom to top; in this order, the chunks alternate between the right and the left of the $V$ (see Figure~\ref{fig:top-level}).

The construction is such that only two consecutive chunks interact.
More precisely, a vertex of a given chunk $T_i$ only sees bits of the terrain contained in $T_{i-1}$, $T_i$, and $T_{i+1}$.
Half-way to the top is the chunk $T_0$ that can be seen as the \emph{initial} one.
It contains the encoding of \emph{all} the variables of the \textsc{Planar 3-SAT}-instance.
Concretely, the reasonable choices to place guards on the chunk $T_0$ are interpreted as setting each variable to either \emph{true} or \emph{false}.
Let $(\mathcal C^+,\mathcal C^-)$ be the bipartition of the clauses into two sets with a removal ordering for the variables ordered as $x_1, x_2, \ldots, x_n$.
Let $C^+_1, C^+_2, \ldots, C^+_s$ (resp.~$C^-_1, C^-_2, \ldots, C^-_{m-s}$) be the order in which the clauses of $\mathcal C^+$ (resp.~$\mathcal C^-$) disappear in this removal ordering.
Every chunk below $T_0$, i.e., with a negative index, are dedicated to checking the clauses of $\mathcal C^-$ in the order $C^-_1, C^-_2, \ldots, C^-_{m-s}$, while every chunk above $T_0$, i.e., with a positive index, will check \emph{if the clauses of $C^+$ are satisfied} in the order $C^+_1, C^+_2, \ldots, C^+_s$.
The placement of the chunks will \emph{propagate downward and upward} the truth assignment of $T_0$, and simulate the operations of a removal ordering: checking/removing a clause and its middle variable, removing a useless variable.
Note that for those gadgets, we will have to distinguish if we are \emph{going up} ($\mathcal C^+$) or \emph{going down} ($\mathcal C^-$).
In addition, the respective position of the positive and negative literals of a variable appearing in the next clause to check will matter.
So, we will require a gadget to invert those two literals if needed.

To sum up, the reduction comprises the following gadgets: encoding a variable (variable gadget), propagation of its assignment from one chunk to a consecutive chunk (interaction of two variable gadgets), inverting its two literals (inverter), checking a clause \emph{upward} and removing the henceforth useless middle variable (upward clause gadget), checking a clause \emph{downward} and removing the henceforth useless middle variable (downward clause gadget), removing a variable (upward/downward deletion gadget).
Although King and Krohn rather crucially rely on having different slopes in the terrain, we will mimic their construction gadget by gadget with an orthogonal terrain.
We start by showing how to simulate a restricted form of a triangular pocket.
This will prove to be a useful building block.

\begin{figure}[h!]
  \centering
  \begin{tikzpicture}[scale=0.8]
  
    \coordinate (v0) at (-0.5,3) {} ;
    \coordinate (v1) at (0,3) {} ;
    \coordinate (v2) at (0,0) {} ;
    \coordinate (v3) at (0.2,0) {} ;
    \coordinate (v4) at (0.2,-1) {} ;
    \coordinate (v5) at (0.4,-1) {} ;
    \coordinate (v6) at (0.4,-3.5) {} ;
    \coordinate (v7) at (0.55,-3.5) {} ;
    \coordinate (v8) at (0.55,-2) {} ;
    \coordinate (v9) at (1,-2) {} ;
    \coordinate (v10) at (1,2) {} ;
    \coordinate (v11) at (1.5,2) {} ;

    \node[vp] at (v2) {} ;
    \node at (-0.15,-0.15) {$a$} ;
    \node[vp] at (v5) {} ;
    \node at (0.25,-1.2) {$u$} ;
    \node[vp] at (v6) {} ;
    \node at (0.25,-3.5) {$p$} ;
    
    \draw (v0) -- (v1) -- (v2) -- (v3) ;
    \draw[very thick] (v3) -- (v4) -- (v5) -- (v6) -- (v7) -- (v8) -- (v9) -- (v10) ;
    \fill[opacity=0.2] (v3) -- (v4) -- (v5) -- (v6) -- (v7) -- (v8) -- (v9) -- (v10) -- cycle ;
    \draw (v10) -- (v11) ;
    \draw[dotted] (v3) -- (v10) ;

    \node at (2.6,0) {$\rightarrow$} ;
    
    \begin{scope}[xshift=4.2cm]
    \coordinate (w0) at (-0.5,3) {} ;
    \coordinate (w1) at (0,3) {} ;
    \coordinate (w2) at (0,0) {} ;
    \coordinate (w3) at (0.2,0) {} ;
    \coordinate (w10) at (1,2) {} ;
    \coordinate (w11) at (1.5,2) {} ;

    \node (tw) at (0.1,-0.2) {$\varepsilon$} ;
    
    \draw (w0) -- (w1) -- (w2) -- (w3) -- (w10) -- (w11) ;

    \node[vp] at (w2) {} ;
    \node at (-0.15,-0.15) {$a$} ;
    \end{scope}
    
    \node at (6.3,0) {$\rightarrow$} ;

    \begin{scope}[xshift=8cm]
    \coordinate (w0) at (-0.5,3) {} ;
    \coordinate (w1) at (0,3) {} ;
    \coordinate (w2) at (0,0) {} ;
    \coordinate (w10) at (1,2) {} ;
    \coordinate (w11) at (1.5,2) {} ;
    
    \draw (w0) -- (w1) -- (w2) -- (w10) -- (w11) ;

    \node[vp] at (w2) {} ;
    \node at (-0.15,-0.15) {$a$} ;
    \end{scope}
    
  \end{tikzpicture}
  \caption{Simulation of a right trapezoid pocket and a right triangular pocket. The right triangular pocket is obtained from the right trapezoid by making the distance $\varepsilon$ sufficiently small.}
  \label{fig:rect-triangular-pocket}
\end{figure}
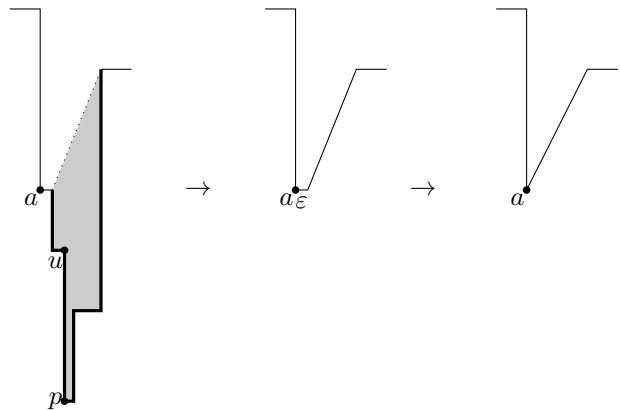

The simulation of a \emph{right trapezoid pocket} giving rise to the \emph{right triangular pocket} is depicted on Figure~\ref{fig:rect-triangular-pocket}.
The idea is that the vertex $p$ at the bottom of the pit is only seen by four vertices (no vertex outside this gadget will be able to see $p$).
Among those four vertices, $u$ sees a strict superset of what the others see.
Hence, we can assume with no loss of generality that a guard is placed on $u$.
Now, $u$ sees the part of the terrain represented in bold.
Even if vertex $u$ sees a part of the vertical edge incident to $a$ (actually the construction could avoid it), this information can be discarded since the guard responsible for seeing $a$ will see this edge entirely.
Everything is therefore equivalent to guarding the terrain with the right trapezoid pocket drawn in the middle of Figure~\ref{fig:rect-triangular-pocket} with a budget of guards decreased by one.
If the length of the horizontal edge incident to $a$ is made small enough, then every guard seeing $a$ will see the whole edge, thereby simulating the right triangular pocket to the right of the figure.

The acute angle made by the right triangular pocket and the altitude of the leftmost and rightmost horizontal edge in this gadget can be set at our convenience.
We will represent triangular pockets in the upcoming gadgets.
The reader should keep in mind that they are actually a shorthand for a rectilinear subterrain. 

\begin{figure}[h!]
  \centering
  \begin{tikzpicture}[scale=0.8]
  
    \coordinate (v0) at (-1.3,3) {} ;
    \coordinate (v1) at (-0.8,3) {} ;

    \coordinate (v2) at (-0.8,-1.5) {} ;
    \coordinate (v3) at (-0.35,-1.5) {} ;
    \coordinate (v4) at (-0.35,-3.5) {} ;
    \coordinate (v5) at (-0.2,-3.5) {} ;
    \coordinate (v6) at (-0.2,-1) {} ;
    \coordinate (v7) at (0,-1) {} ;
     
    \coordinate (v8) at (0,0) {} ;
    \coordinate (v9) at (0.2,0) {} ;
    \coordinate (v10) at (0.2,-1) {} ;
    \coordinate (v11) at (0.4,-1) {} ;
    \coordinate (v12) at (0.4,-3.5) {} ;
    \coordinate (v13) at (0.55,-3.5) {} ;
    \coordinate (v14) at (0.55,-1.5) {} ;
    \coordinate (v15) at (1,-1.5) {} ;
    \coordinate (v16) at (1,2) {} ;
    \coordinate (v17) at (1.5,2) {} ;

    \node[vp] at (v6) {} ;
    \node[vp] at (v11) {} ;
    \node at (0.25,-1.2) {$u$} ;
    \node at (-0.05,-1.2) {$v$} ;
    
    \draw (v0) -- (v1) ;
    \draw [very thick] (v1) -- (v2) -- (v3) --  (v4) -- (v5) -- (v6) -- (v7) -- (v8) ;
    \draw (v8) -- (v9) ;
    \draw[very thick] (v9) -- (v10) -- (v11) -- (v12) -- (v13) -- (v14) -- (v15) -- (v16) ;
    \fill[opacity=0.2] (v9) -- (v10) -- (v11) -- (v12) -- (v13) -- (v14) -- (v15) -- (v16) -- cycle ;
    \fill[opacity=0.2] (v1) -- (v2) -- (v3) -- (v4) -- (v5) -- (v6) -- (v7) -- (v8) -- cycle ;
    \draw (v16) -- (v17) ;
    \draw[dotted] (v1) -- (v8) ;
    \draw[dotted] (v9) -- (v16) ;

    \node at (2.35,0) {$\rightarrow$} ;
    
    \begin{scope}[xshift=4cm]
    \coordinate (w0) at (-1.3,3) {} ;
    \coordinate (w1) at (-0.8,3) {} ;
    \coordinate (w2) at (0,0) {} ;
    \coordinate (w3) at (0.2,0) {} ;
    \coordinate (w10) at (1,2) {} ;
    \coordinate (w11) at (1.5,2) {} ;

    \node (tw) at (0.1,-0.3) {$\varepsilon$} ;
    
    \draw (w0) -- (w1) -- (w2) -- (w3) -- (w10) -- (w11) ;
    \end{scope}
    
    \node at (6.2,0) {$\rightarrow$} ;

    \begin{scope}[xshift=8cm]
    \coordinate (x0) at (-1.3,3) {} ;
    \coordinate (x1) at (-0.8,3) {} ;
    \coordinate (x2) at (0,0) {} ;
    \coordinate (x10) at (1,2) {} ;
    \coordinate (x11) at (1.5,2) {} ;
    
    \draw (x0) -- (x1) -- (x2) -- (x10) -- (x11) ;
    \end{scope}
    
  \end{tikzpicture}
  \caption{Simulation of a trapezoid pocket and a triangular pocket. The triangular pocket is obtained from the trapezoid by making the distance $\varepsilon$ arbitrary small.}
  \label{fig:triangular-pocket}
\end{figure}
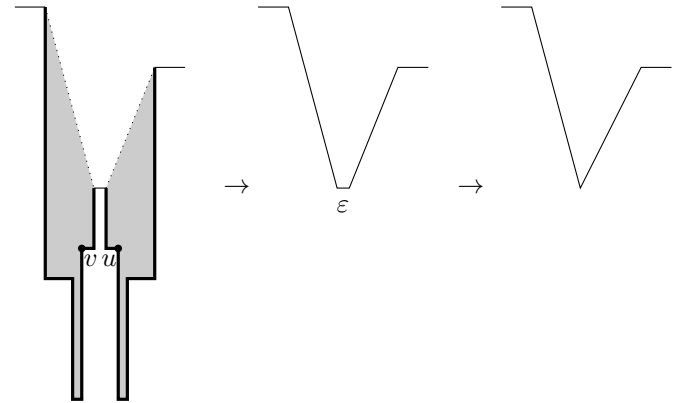

With the same idea, one can simulate a general triangular pocket as depicted on Figure~\ref{fig:triangular-pocket}, with the budget decreased by two guards.
Again, the non-reflex angle made by the triangular pocket and the altitude of the leftmost and rightmost horizontal edges can be freely chosen.
The reason why those triangular pockets do not provide a straightforward reduction from the general \tg problem is that the pocket has to be preceded and succeeded by horizontal edges.

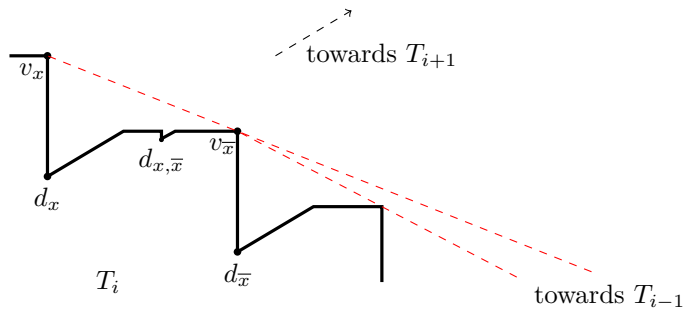
\begin{figure}[h!]
  \centering
  \begin{tikzpicture}
    [extended line/.style={shorten <=-#1},
      extended line/.default=5cm]

    \coordinate (b) at (-0.5,0) {} ;
    \coordinate (x) at (0,0) {} ;
    \coordinate (px) at (0,-1.6) {} ;
    \coordinate (dx) at (1,-1) {} ;
    \coordinate (h1x) at (1.5,-1) {} ;
    \coordinate (h2x) at (1.5,-1.1) {} ;
    \coordinate (h2bx) at (1.5,-1.112) {} ;
    \coordinate (h2tx) at (1.5,-1.4) {} ;
    \coordinate (h3x) at (1.68,-1) {} ;
    \coordinate (y) at (2.5,-1) {} ;
    \coordinate (py) at (2.5,-2.6) {} ;
    \coordinate (dy) at (3.5,-2) {} ;
    \coordinate (e) at (4.4,-2) {} ;
    \coordinate (f) at (4.4,-3) {} ;

    \draw[very thick] (b) -- (x) -- (px) -- (dx) -- (h1x) -- (h2x) -- (h3x) -- (y) -- (py) -- (dy) -- (e) -- (f) ;

    \foreach \i in {x,y,px,py}{
      \node[vp] at (\i) {} ;
    }
    \node[vps] at (h2bx) {} ;

    \node at (h2tx) {$d_{x,\overline x}$} ;

    \node at (-0.2,-0.2) {$v_x$} ;
    \node at (2.3,-1.2) {$v_{\overline x}$} ;

    \node at (0,-1.9) {$d_x$} ;
    \node at (2.5,-2.9) {$d_{\overline x}$} ;

    \node at (0.8,-3) {$T_i$} ;
     \node at (7.4,-3.2) {towards $T_{i-1}$} ;

     \draw[dashed,->] (3,0) -- (4,0.6) ;
     \node at (4.4,0) {towards $T_{i+1}$} ;
     
    \begin{scope}[dashed,red]
    \draw[extended line] (y) -- (x) ;
    \draw[extended line=2cm] (e) -- (y) ;
    \end{scope}

  \end{tikzpicture}
  \caption{A variable gadget. We omit the superscript $i$ on all the labels. Placing a guard at vertex $v_x$ to see $d_x$ corresponds to setting variable $x$ to true, while placing it at vertex $v_{\overline x}$ to see $d_{\overline x}$ corresponds to setting $x$ to false.
    Both $v^{i+1}_x$ and $v^{i+1}_{\overline x}$ of $T_{i+1}$ (not represented on this picture) see $d_{x,\overline x}$ of $T_i$.}
  \label{fig:variable-gadget}
\end{figure}

The variable gadget is depicted on Figure~\ref{fig:variable-gadget}.
It is made of three right triangular pockets.
Placing a guard on $v_x$ (resp.~$v_{\overline x}$) is interpreted as setting the variable $x$ to \emph{true} (resp.~\emph{false}).

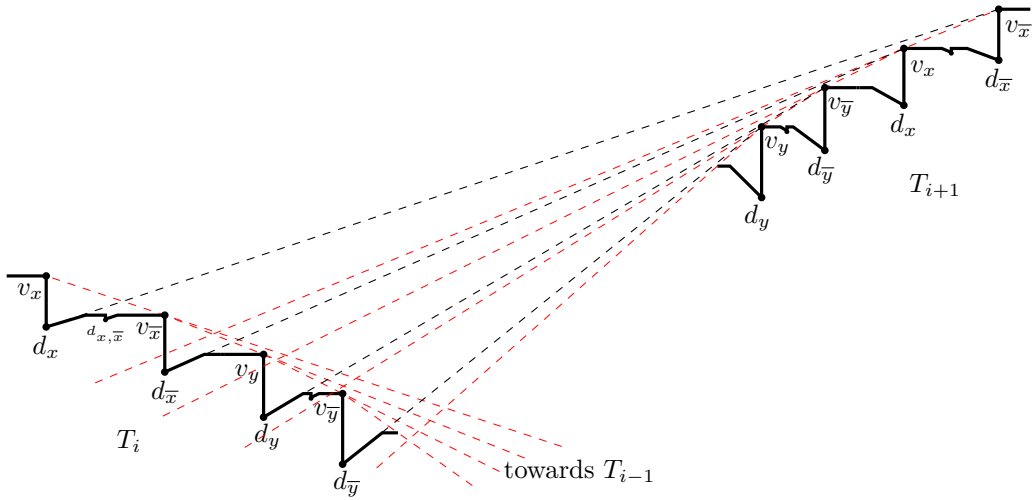
\begin{figure}[h!]
  \centering
  \begin{tikzpicture}
    [scale = 0.52,
      extended line/.style={shorten <= - #1},
      extended line/.default=5cm]

    \def\f{-0.8}

    \foreach \p/\i/\j/\h/\k/\s in {1/0/0/1/-2/5.5,2/25/6.8/\f/-2/-4.4}{

      \begin{scope}[xshift=\i cm, yshift= \j cm]
      
    \coordinate (b1\p) at (-0.5 * \h,0) {} ;
    \coordinate (x1\p) at (0.5 * \h,0) {} ;
    \coordinate (px1\p) at (0.5 * \h,-1.3) {} ;
    \coordinate (dx1\p) at (1.5 * \h,-1) {} ;
    \coordinate (h11\p) at (2 * \h,-1) {} ;
    \coordinate (h21\p) at (2 * \h,-1.1) {} ;
    \coordinate (h2b1\p) at (2 * \h,-1.12) {} ;
    \coordinate (h2t1\p) at (2 * \h,-1.5) {} ;
    \coordinate (h31\p) at (2.3 * \h,-1) {} ;
    \coordinate (y1\p) at (3.5 * \h,-1) {} ;
    \coordinate (py1\p) at (3.5 * \h,-2.45) {} ;
    \coordinate (dy1\p) at (4.5 * \h,-2) {} ;
    \coordinate (e1\p) at (4.9 * \h,-2) {} ;

    \begin{scope}[xshift=\s cm,yshift=\k cm]
    \coordinate (b2\p) at (-0.5 * \h,0) {} ;
    \coordinate (x2\p) at (0.5 * \h,0) {} ;
    \coordinate (px2\p) at (0.5 * \h,-1.6) {} ;
    \coordinate (dx2\p) at (1.5 * \h,-1) {} ;
    \coordinate (h12\p) at (1.7 * \h,-1) {} ;
    \coordinate (h22\p) at (1.7 * \h,-1.1) {} ;
    \coordinate (h2b2\p) at (1.7 * \h,-1.12) {} ;
    \coordinate (h2t2\p) at (1.7 * \h,-1.5) {} ;
    \coordinate (h32\p) at (1.9 * \h,-1) {} ;
    \coordinate (y2\p) at (2.5 * \h,-1) {} ;
    \coordinate (py2\p) at (2.5 * \h,-2.8) {} ;
    \coordinate (dy2\p) at (3.5 * \h,-2) {} ;
    \coordinate (e2\p) at (3.9 * \h,-2) {} ;
    \end{scope}

    \end{scope}

    \begin{scope}[very thick]
    \foreach \i in {1,2}{
      \draw (b\i\p) -- (x\i\p) -- (px\i\p) -- (dx\i\p) -- (h1\i\p) -- (h2\i\p) -- (h3\i\p) -- (y\i\p) -- (py\i\p) -- (dy\i\p) -- (e\i\p) ;
    }
    \draw (e1\p) -- (b2\p) ;
    \end{scope}
    }

    \node at (2.6,-4.2) {$T_i$} ;
    \node at (23,2.3) {$T_{i+1}$} ;

    \node at (14,-5) {towards $T_{i-1}$} ;

    \node at (0.1,-0.4) {$v_x$} ;
    \node at (3.1,-1.4) {$v_{\overline x}$} ;
    \node at (0.5,-1.8) {$d_x$} ;
    \node at (3.5,-2.95) {$d_{\overline x}$} ;

    \node at (h2t11) {\tiny{$d_{x,\overline x}$}} ;
    
    \begin{scope}[xshift=5.6 cm, yshift=-2 cm]
    \node at (0,-0.5) {$v_y$} ;
    \node at (2,-1.5) {$v_{\overline y}$} ;
    \node at (0.5,-2.1) {$d_y$} ;
    \node at (2.5,-3.3) {$d_{\overline y}$} ;
    \end{scope}

    \begin{scope}[xshift=25 cm, yshift=6.8 cm]
    \node at (0.5 * \f + 0.5,-0.5) {$v_{\overline x}$} ;
    \node at (0.5 * \f,-1.8) {$d_{\overline x}$} ;
    \node at (3.5 * \f + 0.5,-1.5) {$v_x$} ;
    \node at (3.5 * \f,-2.95) {$d_x$} ;

    \begin{scope}[xshift=-4.5 cm, yshift=-2 cm]
    \node at (0.5 * \f + 0.5,-0.5) {$v_{\overline y}$} ;
    \node at (2.5 * \f + 0.5,-1.5) {$v_y$} ;
    \node at (0.5 * \f,-2.1) {$d_{\overline y}$} ;
    \node at (2.5 * \f,-3.3) {$d_y$} ;
    \end{scope}

    \end{scope}

    \begin{scope}[dashed, very thin, red]
    \draw[extended line=5.5cm] (y11) -- (x11) ;
    \draw[extended line=3.7cm] (x21) -- (y11) ;
    \draw[extended line=2.3cm] (y21) -- (x21) ;
    \draw[extended line=1.2cm] (e21) -- (y21) ;

    \draw[extended line=11.5cm] (y12) -- (x12) ;
    \draw[extended line=9.7cm] (x22) -- (y12) ;
    \draw[extended line=8cm] (y22) -- (x22) ;
    \draw[extended line=6cm] (e22) -- (y22) ;
    \end{scope}

    \begin{scope}[dashed]
    \draw (x12) -- (px11) ;
    \draw (y12) -- (py11) ;
    \draw (x22) -- (px21) ;
    \draw (y22) -- (py21) ;
    \end{scope}

    \foreach \i/\p in {1/1,1/2,2/1,2/2}{
      \node[vp] at (x\i\p) {} ;
      \node[vp] at (y\i\p) {} ;
      \node[vp] at (px\i\p) {} ;
      \node[vp] at (py\i\p) {} ;
      \node[vps] at (h2b\i\p) {} ;
    }
  \end{tikzpicture}
  \caption{Propagating variable assignments upward and downward.
    Note that the positive literal alternates being above or below the negative literal.
  We represent two variables $x$ and $y$ to illustrate how the corresponding gadgets are not interfering.}
  \label{fig:propagator}
\end{figure}

On Figure~\ref{fig:propagator} is represented the propagation of a variable assignment from one chunk to the next chunk.
On all the upcoming figures, we adopt the convention that red dashed lines materialize a blocked visibility (the vertex cannot see anything \emph{below this line}) and black dashed lines highlight important visibility which sets apart the vertex from other vertices.
Say, one places a guard at vertex $v^i_{\overline x}$ to see (among other things) the vertex $d^i_{\overline x}$.
Now, $d^i_x$ and $d^i_{x,\overline x}$ remain to be seen.
The only way of guarding them with one guard is to place it at vertex $v^{i+1}_{\overline x}$.
Indeed, only vertices on the chunk $T_{i+1}$ can possibly see both.
But the vertices higher than $v^{i+1}_{\overline x}$ cannot see them because their visibility is blocked by $v^{i+1}_{\overline x}$ or a vertex to its right, while the vertices lower than $v^{i+1}_{\overline x}$ are too low to see the very bottom of those two triangular pockets.
The same mechanism (too high $\rightarrow$ blocked visibility, too low $\rightarrow$ too flat angle) is used to ensure that the different variables do not interfere.

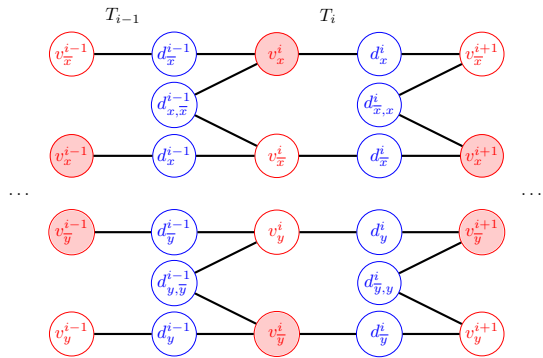
\begin{figure}[h!]
  \centering
  \resizebox{210pt}{!}{
  \begin{tikzpicture}

    \foreach \i in {-5,5}{
      \node at (\i,-0.75) {\dots} ;
    }

    \node at (1,2.75) {$T_i$} ;
    \node at (-3,2.75) {$T_{i-1}$} ;
    
    \foreach \v/\j in {x/0,y/-3.5}{
      \begin{scope}[yshift = \j cm]
    
    \node[draw,circle,red] (vn2\v) at (0,0) {$v^i_{\overline \v}$} ;
    \node[draw,circle,red] (vp2\v) at (0,2) {$v^i_\v$} ;

    \node[draw,circle,red,inner sep=0.03cm] (vn1\v) at (-4,2) {$v^{i-1}_{\overline \v}$} ;
    \node[draw,circle,red,inner sep=0.03cm] (vp1\v) at (-4,0) {$v^{i-1}_\v$} ;

    \node[draw,circle,red,inner sep=0.03cm] (vn3\v) at (4,2) {$v^{i+1}_{\overline \v}$} ;
    \node[draw,circle,red,inner sep=0.03cm] (vp3\v) at (4,0) {$v^{i+1}_\v$} ;

    \node[draw,circle,blue] (dn2\v) at (2,0) {$d^i_{\overline \v}$} ;
    \node[draw,circle,blue,inner sep=0.04cm] (dpn2\v) at (2,1) {$d^i_{\overline \v,\v}$} ;
    \node[draw,circle,blue] (dp2\v) at (2,2) {$d^i_\v$} ;

    \node[draw,circle,blue,inner sep=0.03cm] (dn1\v) at (-2,2) {$d^{i-1}_{\overline \v}$} ;
    \node[draw,circle,blue,inner sep=0.03cm] (dpn1\v) at (-2,1) {$d^{i-1}_{\v,\overline \v}$} ;
    \node[draw,circle,blue,inner sep=0.03cm] (dp1\v) at (-2,0) {$d^{i-1}_\v$} ;

    \foreach \i [count=\j from 2] in {1,2}{
      \begin{scope}[very thick]
      \draw (vn\i\v) -- (dn\i\v) ;
      \draw (vp\i\v) -- (dp\i\v) ;
      \draw (vp\j\v) -- (dpn\i\v) ;
      \draw (vp\j\v) -- (dn\i\v) ;
      \draw (vn\j\v) -- (dpn\i\v) ;
      \draw (vn\j\v) -- (dp\i\v) ;
      \end{scope}
    }

      \end{scope}
    }
    \foreach \i in {vp1x,vp2x,vp3x,vn1y,vn2y,vn3y}{
      \node[circle,fill=red,opacity=0.2,inner sep=-0.3cm] at (\i) {};
    }
  \end{tikzpicture}
  }
  \caption{The red-blue domination graph for variable-assignment propagation.
  }
  \label{fig:graph-propagator}
\end{figure}

Symmetrically, the only vertex seeing both $d^i_{x,\overline x}$ and $d^i_{\overline x}$ is $v^{i+1}_x$.
So, placing a guard at $v^i_x$ forces to place the other guard at $v^{i+1}_x$.
Observe that the chosen literal goes from being above (resp.~below) in chunk $T_i$ to being below (resp.~above) in chunk $T_{i+1}$.
Also, each $d$-vertex (i.e., vertex of the form $d^{\bullet}_{\bullet}$) has its visibility contained in the one of a $v$-vertex (of the form $v^{\bullet}_{\bullet}$).
Actually, each non $v$-vertex has its visibility contained in the one of a $v$-vertex.
Furthermore, seeing the $d$-vertices with $v$-vertices is enough to see the entire subterrain/chunk. 
Hence, the problem can be seen as a red-blue domination: taking $v$-vertices (red) to dominate the $d$-vertices (blue).
The red-blue visibility graph corresponding to the propagation of variable assignments is represented on Figure~\ref{fig:graph-propagator}.
It can be observed that the only way of guarding the $3z$ $d$-vertices on chunk $T^i$ (corresponding to $z$ vertices) with a budget of $2z$ guards is to place $z$ guards on $v$-vertices of chunk $T_i$ and $z$ guards on $v$-vertices of chunk $T_{i+1}$ in a consistent way: the assignment of each variable is preserved. 

We also need an alternative way of propagating truth assignments such that the chosen literal stays above or stays below on its respective chunk.
This gadget is called \emph{inverter}.
It requires an extra guard compared to the usual propagation.
The inverter gadget allows us to position the three literals of the clause to check and delete at the right spots.

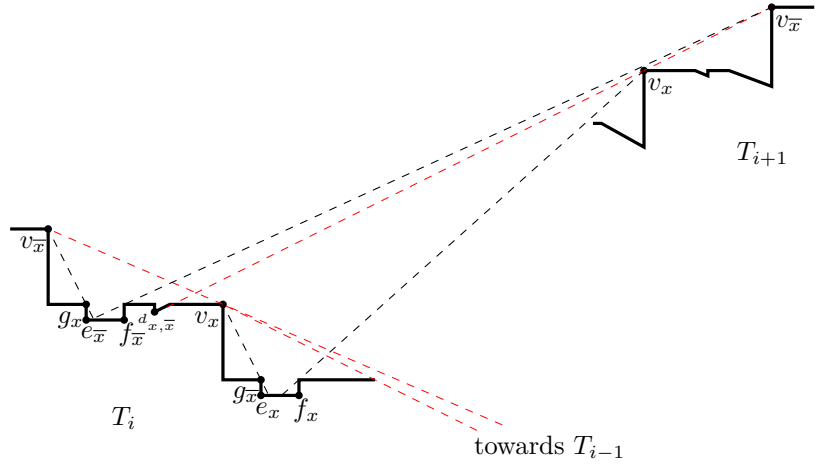
\begin{figure}[h!]
  \centering
  \begin{tikzpicture}
    [extended line/.style={shorten <= - #1},
      extended line/.default=5cm]
    \def\h{-0.8}

    \coordinate (i1) at (-0.5,0) {};
    \coordinate (i2) at (0,0) {};
    \coordinate (ti2) at (-0.2,-0.2) {};
    \coordinate (i3) at (0,-1) {};
    \coordinate (i4) at (0.5,-1) {};
    \coordinate (ti4) at (0.3,-1.2) {};
    \coordinate (i5) at (0.5,-1.207) {};
    \coordinate (ti5) at (0.6,-1.4) {};
    \coordinate (i6) at (1,-1.207) {};
    \coordinate (ti6) at (1.1,-1.4) {};
    \coordinate (i7) at (1,-1) {};

    \coordinate (i7a) at (1.4,-1) {};
    \coordinate (i7b) at (1.4,-1.1) {};
    \coordinate (ti7b) at (1.43,-1.25) {};
    \coordinate (i7c) at (1.6,-1) {};

    \coordinate (i8) at (2.3,-1) {};
    \coordinate (ti8) at (2.1,-1.2) {};
    \coordinate (i9) at (2.3,-2) {};
    \coordinate (i10) at (2.8,-2) {};
    \coordinate (ti10) at (2.6,-2.2) {};
    \coordinate (i11) at (2.8,-2.207) {};
    \coordinate (ti11) at (2.9,-2.4) {};
    \coordinate (i12) at (3.3,-2.207) {};
    \coordinate (ti12) at (3.4,-2.4) {};
    \coordinate (i13) at (3.3,-2) {};
    \coordinate (i14) at (4.3,-2) {};
    
    \draw[very thick] (i1) -- (i2) -- (i3) -- (i4) -- (i5) -- (i6) -- (i7) -- (i7a) --(i7b) --(i7c) -- (i8) -- (i9) -- (i10) --  (i11) -- (i12) -- (i13) -- (i14) ;

    \foreach \i/\j in {14/4}{
      \begin{scope}[scale=0.7,xshift=\i cm, yshift= \j cm]
    \coordinate (b1) at (-0.5 * \h,0.2) {} ;
    \coordinate (x1) at (0.5 * \h,0.2) {} ;
    \coordinate (tx1) at (0.1 * \h,-0.1) {} ;
    \coordinate (px1) at (0.5 * \h,-1.3) {} ;
    \coordinate (dx1) at (1.5 * \h,-1) {} ;
    \coordinate (h11) at (2 * \h,-1) {} ;
    \coordinate (h21) at (2 * \h,-1.1) {} ;
    \coordinate (h2b1) at (2 * \h,-1.12) {} ;
    \coordinate (h2t1) at (2 * \h,-1.5) {} ;
    \coordinate (h31) at (2.3 * \h,-1) {} ;
    \coordinate (y1) at (3.5 * \h,-1) {} ;
    \coordinate (ty1) at (3.1 * \h,-1.3) {} ;
    \coordinate (py1) at (3.5 * \h,-2.45) {} ;
    \coordinate (dy1) at (4.5 * \h,-2) {} ;
    \coordinate (e1) at (4.7 * \h,-2) {} ;
      \end{scope}

      \begin{scope}[very thick]
        \draw (b1) -- (x1) -- (px1) -- (dx1) -- (h11) -- (h21) -- (h31) -- (y1) -- (py1) -- (dy1) -- (e1) ;
      \end{scope}
    }

    \foreach \v in {i2,i4,i5,i6,i8,i10,i11,i12,x1,y1,i7b}{
      \node[vp] at (\v) {} ;
    }

    \node at (ty1) {$v_x$} ;
    \node at (tx1) {$v_{\overline x}$} ;
    \node at (ti8) {$v_x$} ;
    \node at (ti2) {$v_{\overline x}$} ;

    \node at (ti10) {$g_{\overline x}$} ;
    \node at (ti4) {$g_x$} ;

    \node at (ti7b) {\tiny{$d_{x,\overline x}$}} ;

    \node at (ti5) {$e_{\overline x}$} ;
    \node at (ti6) {$f_{\overline x}$} ;

    \node at (ti11) {$e_x$} ;
    \node at (ti12) {$f_x$} ;

    \node at (1,-2.5) {$T_i$} ;
    \node at (6.6,-2.9) {towards $T_{i-1}$} ;
    \node at (9.4,1) {$T_{i+1}$} ;
    
    \begin{scope}[dashed, very thin, red]
      \draw[extended line=4cm] (i8) -- (i2) ;
      \draw[extended line=1.5cm] (i14) -- (i8) ;

      \draw[extended line=7cm] (y1) -- (x1) ;
    \end{scope}

     \begin{scope}[dashed, thin]
      \draw[extended line=0.2cm] (i4) -- (i2) ;
      \draw[extended line=0.2cm] (i10) -- (i8) ;

      \draw[extended line=0.3cm] (i13) -- (y1) ;
      \draw[extended line=0.4cm] (i7) -- (x1) ;
     \end{scope}
    
  \end{tikzpicture}
  \caption{The inverter gadget. We omit the superscripts $i$ and $i+1$.
  If a guard should be placed on at least one vertex among $v_{\overline x}^\ell$ and $v_x^\ell$ (for $\ell \in \{i,i+1\}$), then the two ways of seeing the four vertices $e_{\overline x}^i$, $f_{\overline x}^i$,  $e_x^i$,  $f_x^i$ with three guards are $\{v_{\overline x}^i, g_{\overline x}^i, v_{\overline x}^{i+1}\}$ and $\{v_x^i,g_x^i, v_x^{i+1}\}$.}
  \label{fig:inverter-gadget}
\end{figure}

It consists of a right triangular pocket whose bottom vertex is $d^i_{x,\overline x}$ surrounded by two rectangular pockets whose bottom vertices $e^i_x, f^i_x$ and $e^i_{\overline x}, f^i_{\overline x}$ are only seen among the $v$-vertices by $v^{i+1}_x, v^i_x$ and $v^{i+1}_{\overline x}, v^i_{\overline x}$, respectively.
On top of the rectangular pockets, $g^i_x$ sees both $e^i_{\overline x}$ and $f^i_{\overline x}$, whereas $g^i_{\overline x}$ sees both $e^i_x$ and $f^i_x$.
Actually, $g^i_\ell$ is only one of the four vertices seeing both $e^i_\ell$ and $f^i_\ell$ (which includes $e^i_\ell$ and $f^i_\ell$ themselves).
We choose $g^i_\ell$ as a representative of this class.
What matters to us is that the four vertices seeing both $e^i_\ell$ and $f^i_\ell$ do not see anything more than the rectangular pocket; the other parts of the terrain that they might guard are seen by any $v$-vertex on chunk $T_{i+1}$ anyway.

\begin{figure}[h!]
  \centering
  \resizebox{100pt}{!}{
  \begin{tikzpicture}

    \node at (1,3.6) {$T_i$} ;
    
    \node[draw,circle,red] (vp1) at (0,-1) {$v^i_x$} ;
    \node[draw,circle,red] (ng1) at (0,0) {$g^i_{\overline x}$} ;
    \node[draw,circle,red] (g1) at (0,2) {$g^i_x$} ;
    \node[draw,circle,red] (vn1) at (0,3) {$v^i_{\overline x}$} ;

    \node[draw,circle,blue] (f1) at (2,-1) {$f^i_x$} ;
    \node[draw,circle,blue] (e1) at (2,0) {$e^i_x$} ;
    \node[draw,circle,blue,inner sep=0.033cm] (d1) at (2,1) {$d^i_{x,\overline x}$} ;
    \node[draw,circle,blue] (ne1) at (2,2) {$e^i_{\overline x}$} ;
    \node[draw,circle,blue] (nf1) at (2,3) {$f^i_{\overline x}$} ;

    \node[draw,circle,red,inner sep=0.03cm] (vp2) at (4,0) {$v^{i+1}_x$} ;
    \node[draw,circle,red,inner sep=0.03cm] (vn2) at (4,2) {$v^{i+1}_{\overline x}$} ;

    \begin{scope}[very thick]
      \draw (vp2) -- (d1) -- (vn2) ;
      \draw (e1) -- (ng1) -- (f1) ;
      \draw (ne1) -- (g1) -- (nf1) ;
      \draw (vp1) -- (f1) ;
      \draw (vn1) -- (nf1) ;
      \draw (vp2) -- (e1) ;
      \draw (vn2) -- (ne1) ;
    \end{scope}
    
    \foreach \i in {ng1,vn1,vn2}{
      \node[circle,fill=red,opacity=0.2,inner sep=-0.3cm] at (\i) {};
    }
  \end{tikzpicture}
  }
  \caption{The red-blue domination graph for the inverter gadget.
  }
  \label{fig:graph-inverter}
\end{figure}
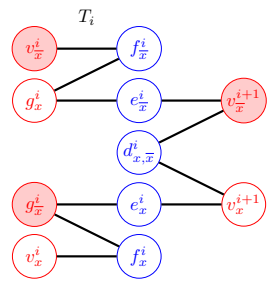

The pockets are designed so that $v^i_x$ and $v^{i+1}_x$ (resp.~$v^i_{\overline x}$ and $v^{i+1}_{\overline x}$) together see the whole edge $e^i_xf^i_x$ (resp.~$e^i_{\overline x}f^i_{\overline x}$) and therefore the entire pocket.
Again, the only two $v$-vertices to see $d^i_{x,\overline x}$ are $v^{i+1}_x$ and $v^{i+1}_{\overline x}$.
The $e$- and $f$-vertices are added to the blue vertices and the $g$-vertices are added to the red vertices, since the latter sees more than the former, and since seeing the $e$- and $f$-vertices are sufficient to also see the $g$-vertices.
The red-blue domination graph is depicted on Figure~\ref{fig:graph-inverter}.

Guarding $d^{i-1}_{x,\overline x}$ (resp.~guarding $d^i_{x,\overline x}$) requires to take one $v$-vertex among $v^i_x, v^i_{\overline x}$ (resp.~$v^{i+1}_x,$ $v^{i+1}_{\overline x}$).
The two only ways of seeing both rectangular pockets with an extra guard is then to place the three guards at $v^i_x, g^i_x, v^{i+1}_x$ or $v^i_{\overline x}, g^i_{\overline x}, v^{i+1}_{\overline x}$; hence the propagation of the truth assignment. 

So far, the gadgets that we presented can be used \emph{going up} along the chunks of positive index as well as \emph{going down} along the chunks of negative index.
For the clause gadgets, we will have to distinguish the \emph{downward clause gadget} when we are below $T_0$ (and going down) and the \emph{upward clause gadget} when we are above $T_0$ (and going down).
The reason we cannot design a single gadget for both situations is that the middle variable which needs be deleted is in one case, in the lower chunk, and in the other case, in the higher chunk.

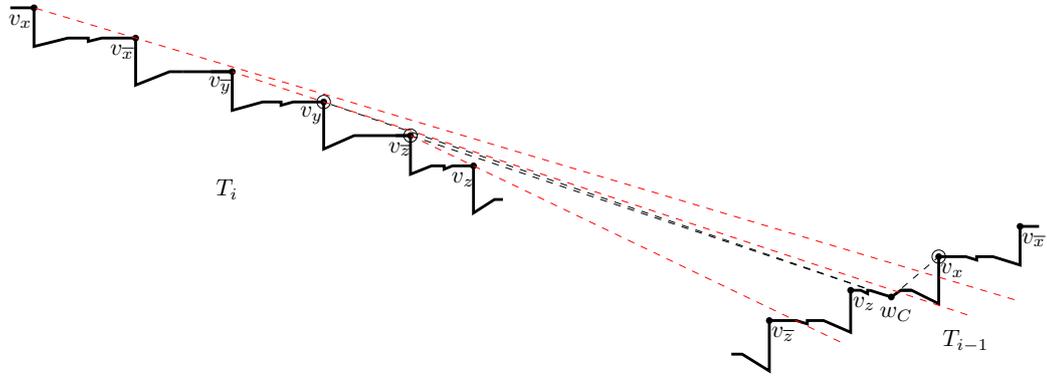
\begin{figure}[h!]
  \centering
  \resizebox{400pt}{!}{
  \begin{tikzpicture}
    [scale = 0.5,
      extended line/.style={shorten <= - #1},
      extended line/.default=5cm]

    \foreach \i/\j/\h/\d/\p in {-1/0/1/0/1,4.9/-1.9/0.9/0/2,10.3/-3.8/0.62/0/3,24/-8.4/-0.8/-0.1/4,29/-6.5/-0.8/0/5}{

      \begin{scope}[xshift=\i cm, yshift=\j cm]
      
    \coordinate (b\p) at (-0.2 * \h,-0.1) {} ;
    \coordinate (tb\p) at (-0.9 * \h,-0.3) {} ;
    \coordinate (x\p) at (0.5 * \h,-0.1) {} ;
    \coordinate (tx\p) at (0.1 * \h,-0.5) {} ;
    \coordinate (ttx\p) at (0 * \h,-0.5) {} ;
    \coordinate (dx\p) at (0.5 * \h,-1.25+\d) {} ;
    \coordinate (px\p) at (1.5 * \h,-1) {} ;
    \coordinate (h1x\p) at (2.1 * \h,-1) {} ;
    \coordinate (h2x\p) at (2.1 * \h,-1.1) {} ;
    \coordinate (h3x\p) at (2.5 * \h,-1) {} ;
    \coordinate (nx\p) at (3.5 * \h,-1) {} ;
    \coordinate (tnx\p) at (3.1 * \h,-1.4) {} ;
    \coordinate (ttnx\p) at (3 * \h,-1.4) {} ;
    \coordinate (ndx\p) at (3.5 * \h,-2.4+\d) {} ;
    \coordinate (npx\p) at (4.5 * \h,-2) {} ;
    \coordinate (f\p) at (5.85 * \h,-2) {} ;
    \coordinate (e\p) at (4.9 * \h,-2) {} ;

      \end{scope}

      \draw[very thick] (x\p) -- (dx\p) -- (px\p) -- (h1x\p) -- (h2x\p) -- (h3x\p) -- (nx\p) -- (ndx\p) -- (npx\p) -- (e\p) ;
    }

    \foreach \v in {x1,x2,x3,x4,x5,nx1,nx2,nx3,nx4,nx5}{
      \node[vp] at (\v) {} ;
    }

    \foreach \v in {nx2,x3,nx5}{
      \node[draw, circle,inner sep=-0.07cm] at (\v) {} ;
    }
    
    \foreach \i in {1,2,3,5}{
      \draw[very thick] (b\i) -- (x\i) ;
    }

    \coordinate (wC) at (24.8,-8.7) ;
    \coordinate (twC) at (24.95,-9.2) ;
    \node[vp] at (wC) {} ;
    \node at (twC) {$w_C$} ;

    \begin{scope}[dashed, thin]
      \draw (wC) -- (nx5) ;
      \draw (wC) -- (nx2) ;
      \draw (wC) -- (x3) ; 
    \end{scope}

    \begin{scope}[very thick]
      \draw (e1) -- (b2) -- (x2) ;
      \draw (e2) -- (b3) -- (x3) ;
      \draw (npx5) -- (e5) -- (wC) -- (b4) ;
      \draw (x4) --++(0.3,0)--++(0.2,-0.1)--++(0,0.1) -- (b4) ;
    \end{scope}

   \node at (tx1) {$v_x$} ;
    \node at (tx2) {$v_{\overline y}$} ;
    \node at (ttx3) {$v_{\overline z}$} ;
    \node at (ttx4) {$v_z$} ;
    \node at (ttx5) {$v_{\overline x}$} ;
    \node at (tnx1) {$v_{\overline x}$} ;
    \node at (tnx2) {$v_y$} ;
    \node at (ttnx3) {$v_z$} ;
    \node at (ttnx4) {$v_{\overline z}$} ;
    \node at (ttnx5) {$v_x$} ;

    \begin{scope}[dashed,very thin,red]
      \draw[extended line=13.5cm] (nx1) -- (x1) ;
      \draw[extended line=10cm] (nx2) -- (x2) ;
      \draw[extended line=6cm] (nx3) -- (x3) ;
    \end{scope}
    
    \node at (5.2,-5.5) {$T_i$} ;
    \node at (27,-10) {$T_{i-1}$} ;
    
  \end{tikzpicture}
  }
  \caption{The downward clause gadget for $C=x \lor y \lor \neg z$.
    We use the usual propagation for variables $x$ and $z$.
    The variable $y$ disappears from $T_{i-1}$ and downward.
    The inverters have been used to place, on $T_i$, the literals of $C$ at positions 1, 4, and 5.
    The vertex $w_C$ is seen only by $v^i_{y}$, $v^i_{\overline z}$, and $v^{i-1}_x$ (circled);
    hence it is seen if and only if the chosen assignment satisfies $C$.}
  \label{fig:downward-clause}
\end{figure}

To check a clause downward on three consecutive variables $x, y, z$, we place on chunk $T_i$, thanks to a preliminary use of inverter gadgets, the three literals satisfying the clause at the relative positions $1$, $4$, and $5$ when the six literals of $x, y, z$ are read from top to bottom.
Figure~\ref{fig:downward-clause} shows the downward clause gadget for the clause $x \lor y \lor \neg z$.
On the chunk $T_{i-1}$ just below, we find the usual encoding of variables $x$ and $z$, which propagates the truth assignment of those two variables.
The variable gadget of $y$ is replaced by the right triangular pocket whose bottom is $d^{i-1}_{y,\overline y}$, and a general triangular pocket whose bottom $w_C$ is only seen among the $v$-vertices by $v_{\ell_1}^{i-1}$ (on chunk $T_{i-1}$) and $v_{\ell_2}^i$ and  $v_{\ell_3}^i$ (on chunk $T_i$), where $C = \ell_1 \lor \ell_2 \lor \ell_3$.
On chunk $T_{i-1}$ and below, no $v$-vertex corresponding to variable $y$ can be found.

\begin{figure}[h!]
  \centering
  \resizebox{100pt}{!}{
  \begin{tikzpicture}

    \node at (0,5.8) {$T_i$} ;
    \node at (-4,5.8) {$T_{i-1}$} ;
    
    \node[draw,circle,red] (z1) at (0,0) {$v^i_z$} ;
    \node[draw,circle,red] (nz1) at (0,1) {$v^i_{\overline z}$} ;
    \node[draw,circle,red] (y1) at (0,2) {$v^i_y$} ;
    \node[draw,circle,red] (ny1) at (0,3) {$v^i_{\overline y}$} ;
    \node[draw,circle,red] (x1) at (0,4) {$v^i_x$} ;
    \node[draw,circle,red] (nx1) at (0,5) {$v^i_{\overline x}$} ;

    \node[draw,circle,blue] (wC) at (-2,2) {$w_C$} ;
    \node[draw,circle,blue,inner sep=0.03cm] (d) at (-2,3) {$d^{i-1}_{y,\overline y}$} ;

    \node[draw,circle,red,inner sep=0.03cm] (z2) at (-4,0) {$v^{i-1}_z$} ;
    \node[draw,circle,red,inner sep=0.03cm] (nz2) at (-4,1) {$v^{i-1}_{\overline z}$} ;
    \node[draw,circle,red,inner sep=0.03cm] (x2) at (-4,4) {$v^{i-1}_x$} ;
    \node[draw,circle,red,inner sep=0.03cm] (nx2) at (-4,5) {$v^{i-1}_{\overline x}$} ;

    \begin{scope}[very thick]
      \draw (y1) -- (d) -- (ny1) ;
      \draw (y1) -- (wC) -- (x2) ;
      \draw (wC) -- (nz1) ;
    \end{scope}
    
    \foreach \i in {nx1,nx2,y1,z1,z2}{
      \node[circle,fill=red,opacity=0.2,inner sep=-0.3cm] at (\i) {};
    }

    \foreach \i/\j in {x1/x2,nx1/nx2,z1/z2,nz1/nz2}{
      \draw[<->] (\i) -- (\j) ;
    }
  \end{tikzpicture}
  }
  \caption{The red-blue domination graph for the downward clause gadget for $C=x \lor y \lor \neg z$.
  The double arcs symbolize that, due to the propagator, the variable-assignment of $x$ and $z$ should be the same between $T_i$ and $T_{i-1}$. The only assignment that does not dominate $w_C$ is $\overline x$, $\overline y$, $z$, as it should.}
  \label{fig:graph-downward}
\end{figure}
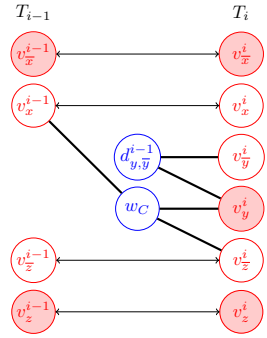

Hence, the vertex $w_C$ is only guarded if the choices of the guards at the $v$-vertices correspond to an assignment satisfying $C$.
The vertex $w_C$ has its visibility contained in the one of a $v$-vertex, hence it is a blue vertex.
The red-blue domination graph associated to a downward clause is represented on Figure~\ref{fig:graph-downward}.

\begin{figure}[h!]
  \centering
  \resizebox{400pt}{!}{
  \begin{tikzpicture}
    [scale = 0.5,
      extended line/.style={shorten <= - #1},
      extended line/.default=5cm]

    \foreach \i/\j/\h/\d/\p in {-1/0/1/0/1,10/-3.2/0.5/0.15/3,25/-1.5/-0.8/0/4,31/5/-0.8/0/5}{

      \begin{scope}[xshift=\i cm, yshift=\j cm]
      
    \coordinate (b\p) at (-0.5 * \h,-0.5) {} ;
    \coordinate (tb\p) at (-0.9 * \h,-0.7) {} ;
    \coordinate (x\p) at (0.5 * \h,-0.5) {} ;
    \coordinate (tx\p) at (0.1 * \h,-0.9) {} ;
    \coordinate (ttx\p) at (0 * \h,-0.9) {} ;
    \coordinate (tttx\p) at (-0.2 * \h,-0.9) {} ;
    \coordinate (dx\p) at (0.5 * \h,-1.3+\d) {} ;
    \coordinate (px\p) at (1.5 * \h,-1) {} ;
    \coordinate (h1x\p) at (2.1 * \h,-1) {} ;
    \coordinate (h2x\p) at (2.1 * \h,-1.1+0.3 * \d) {} ;
    \coordinate (h3x\p) at (2.5 * \h,-1) {} ;
    \coordinate (nx\p) at (3.5 * \h,-1) {} ;
    \coordinate (tnx\p) at (3.1 * \h,-1.4) {} ;
    \coordinate (ttnx\p) at (3 * \h,-1.4) {} ;
    \coordinate (tttnx\p) at (2.8 * \h,-1.4) {} ;
    \coordinate (ndx\p) at (3.5 * \h,-2.3+\d) {} ;
    \coordinate (npx\p) at (4.5 * \h,-2) {} ;
    \coordinate (f\p) at (5.85 * \h,-2) {} ;
    \coordinate (e\p) at (4.9 * \h,-2) {} ;

      \end{scope}

      \draw[very thick] (b\p) -- (x\p) -- (dx\p) -- (px\p) -- (h1x\p) -- (h2x\p) -- (h3x\p) -- (nx\p) -- (ndx\p) -- (npx\p) -- (e\p) ;
    }

    \foreach \i/\j/\h/\p in {5.3/-2.4/0.9/2}{
      \begin{scope}[xshift=\i cm, yshift=\j cm]
    \coordinate (b\p) at (-0.5 * \h,0) {} ;
    \coordinate (tb\p) at (-0.9 * \h,-0.3) {} ;
    \coordinate (x\p) at (1.8 * \h,0) {} ;
    \coordinate (tx\p) at (1.3 * \h,-0.4) {} ;
    \coordinate (dx\p) at (1.8 * \h,-0.7) {} ;
    \coordinate (nx\p) at (3.8 * \h,-0.7) {} ;
    \coordinate (tnx\p) at (3.4 * \h,-1.1) {} ;
    \coordinate (f\p) at (3.8 * \h,-1.3) {} ;
    \coordinate (e\p) at (5 * \h,-1.3) {} ;
      \end{scope}
    }

    \draw[very thick] (b2) -- (x2) -- (dx2) -- (nx2) -- (f2) -- (e2) ;

    \draw[very thick] (e1) -- (f1) -- (b2) ;
    \draw[very thick] (npx5) -- (e5) -- (25.4,3) -- (b4) -- (x4) ;

    \foreach \v in {x1,x2,x3,x4,x5,nx1,nx2,nx3,nx4,nx5,b2}{
      \node[vp] at (\v) {} ;
    }

    \foreach \v in {x2,x4,nx5}{
      \node[draw, circle,inner sep=-0.07cm] at (\v) {} ;
    }

    \node at (tx1) {$v_x$} ;
    \node at (tx2) {$v_{\overline y}$} ;
    \node at (tttx3) {$v_{\overline z}$} ;
    \node at (ttx4) {$v_z$} ;
    \node at (ttx5) {$v_{\overline x}$} ;
    \node at (tnx1) {$v_{\overline x}$} ;
    \node at (tnx2) {$v_y$} ;
    \node at (tttnx3) {$v_z$} ;
    \node at (ttnx4) {$v_{\overline z}$} ;
    \node at (ttnx5) {$v_x$} ;

    \node at (tb2) {$w_C$} ;

    \begin{scope}[dashed,thin]
      \draw (b2) -- (x4) ;
      \draw (b2) -- (nx5) ; 
    \end{scope}

    \node at (5.2,-5.5) {$T_i$} ;
    \node at (29,0) {$T_{i+1}$} ;

    \begin{scope}[dashed,very thin,red]
      \draw[extended line=15cm] (nx5) -- (x5) ;
      \draw[extended line=7cm] (nx4) -- (x4) ;
    \end{scope}
    
  \end{tikzpicture}
  }
  \caption{The upward clause gadget for $C=x \lor \neg y \lor z$.
    We use the usual propagation for variables $x$ and $z$.
    The variable $y$ disappears from $T_{i+1}$ and upward.
    The inverters have been used to place, on $T_i$, the literals of $C$ at positions 1, 3, and 6.
    The vertex $w_C$ is seen only by $v^i_{\overline y}$, $v^{i+1}_x$, and $v^{i+1}_z$ (circled);
    hence it is seen if and only if the chosen assignment satisfies $C$.}
  \label{fig:upward-clause}
\end{figure}
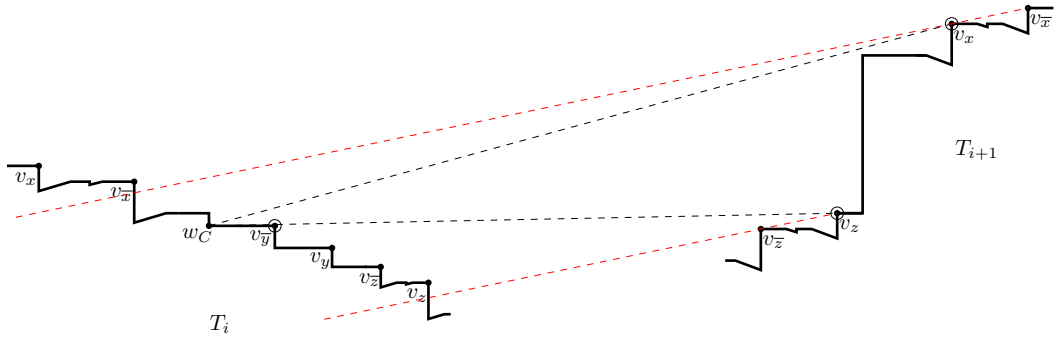

To check a clause upward on three consecutive variables $x, y, z$, we place on chunk $T_i$, thanks to a preliminary use of inverter gadgets, the three literals satisfying the clause at the relative positions $1$, $3$, and $6$ when the six literals of $x, y, z$ are read from top to bottom.
We exclude the three right triangular pockets for the encoding of the middle variable $y$.
At the same altitude as the $v$-vertex corresponding to the literal of $y$ satisfying the clause, we have a designated vertex $w_C$.
On the chunk $T_{i+1}$, we find the usual encoding of variables $x$ and $z$, which propagates the truth assignment of those two variables, but the encoding of variable $y$ is no longer present (in this chunk and in all the chunks above).
Figure~\ref{fig:upward-clause} shows the upward clause gadget for the clause $x \lor \neg y \lor z$.

\begin{figure}[h!]
  \centering
  \resizebox{100pt}{!}{
  \begin{tikzpicture}

    \node at (0,5.8) {$T_i$} ;
    \node at (4,5.8) {$T_{i+1}$} ;
    
    \node[draw,circle,red] (z1) at (0,0) {$v^i_z$} ;
    \node[draw,circle,red] (nz1) at (0,1) {$v^i_{\overline z}$} ;
    \node[draw,circle,red] (y1) at (0,2) {$v^i_y$} ;
    \node[draw,circle,red] (ny1) at (0,3) {$v^i_{\overline y}$} ;
    \node[draw,circle,red] (x1) at (0,4) {$v^i_x$} ;
    \node[draw,circle,red] (nx1) at (0,5) {$v^i_{\overline x}$} ;

    \node[draw,circle,blue] (wC) at (2,2.5) {$w_C$} ;

    \node[draw,circle,red,inner sep=0.03cm] (z2) at (4,0) {$v^{i+1}_z$} ;
    \node[draw,circle,red,inner sep=0.03cm] (nz2) at (4,1) {$v^{i+1}_{\overline z}$} ;
    \node[draw,circle,red,inner sep=0.03cm] (x2) at (4,4) {$v^{i+1}_x$} ;
    \node[draw,circle,red,inner sep=0.03cm] (nx2) at (4,5) {$v^{i+1}_{\overline x}$} ;

    \begin{scope}[very thick]
      \draw (ny1) -- (wC) -- (x2) ;
      \draw (wC) -- (z2) ;
    \end{scope}
    
    \foreach \i in {x1,x2,y1,nz1,nz2}{
      \node[circle,fill=red,opacity=0.2,inner sep=-0.3cm] at (\i) {};
    }

    \foreach \i/\j in {x1/x2,nx1/nx2,z1/z2,nz1/nz2}{
      \draw[<->] (\i) -- (\j) ;
    }
  \end{tikzpicture}
  }
  \caption{The red-blue domination graph for the upward clause gadget for $C=x \lor \neg y \lor z$.
  The double arcs symbolize that, due to the propagator, the variable-assignment of $x$ and $z$ should be the same between $T_i$ and $T_{i+1}$. The only assignment that does not dominate $w_C$ is $\overline x$, $y$, $\overline z$, as it should.}
  \label{fig:graph-upward}
\end{figure}
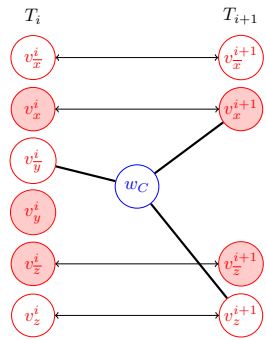

The vertex $w_C$ is only seen among the $v$-vertices by $v_{\ell_2}^i$ (on chunk $T_i$) and $v_{\ell_1}^{i+1}$ and $v_{\ell_3}^{i+1}$ (on chunk $T_{i+1}$), where $C = \ell_1 \lor \ell_2 \lor \ell_3$.
The particularity of two consecutive chunks encoding an upward clause gadget is that $T_i$ is not entirely below $T_{i+1}$.
In fact, all the encodings of variables above $y$ on chunk $T_{i+1}$ are above all the encodings of variables above $y$ on chunk $T_i$.
The latter are above all the encodings of variables below $y$ on chunk $T_{i+1}$, which are, in turn, above all the encodings of variables below $y$ on chunk $T_i$.
Again, the vertex $w_C$ is only guarded if the choices of the guards at the $v$-vertices correspond to an assignment satisfying $C$, as depicted in Figure~\ref{fig:graph-upward}.

Finally, we design variable deletion gadgets.
Recall that we sometimes need to remove a variable which does not appear in any clauses anymore (and was never a middle variable).
As for clause gadgets, we have to distinguish \emph{downward deletion gadget} and \emph{upward deletion gadget}.
Both gadgets can be thought as a simplification of the corresponding clause gadget where we flatten the region which should normally contain $w_C$. 

\begin{figure}[h!]
  \centering
  \begin{tikzpicture}
    [scale = 0.52,
      extended line/.style={shorten <= - #1},
      extended line/.default=5cm]

    \def\f{-0.8}

    \foreach \p/\i/\j/\h/\k/\s in {1/0/0/1/-2/5.5}{

      \begin{scope}[xshift=\i cm, yshift= \j cm]
      
    \coordinate (b1\p) at (-0.5 * \h,0) {} ;
    \coordinate (x1\p) at (0.5 * \h,0) {} ;
    \coordinate (px1\p) at (0.5 * \h,-1.3) {} ;
    \coordinate (dx1\p) at (1.5 * \h,-1) {} ;
    \coordinate (h11\p) at (2 * \h,-1) {} ;
    \coordinate (h21\p) at (2 * \h,-1.1) {} ;
    \coordinate (h2b1\p) at (2 * \h,-1.12) {} ;
    \coordinate (h2t1\p) at (2 * \h,-1.5) {} ;
    \coordinate (h31\p) at (2.3 * \h,-1) {} ;
    \coordinate (y1\p) at (3.5 * \h,-1) {} ;
    \coordinate (py1\p) at (3.5 * \h,-2.45) {} ;
    \coordinate (dy1\p) at (4.5 * \h,-2) {} ;
    \coordinate (e1\p) at (4.9 * \h,-2) {} ;

    \begin{scope}[xshift=\s cm,yshift=\k cm]
    \coordinate (b2\p) at (-0.5 * \h,0) {} ;
    \coordinate (x2\p) at (0.5 * \h,0) {} ;
    \coordinate (px2\p) at (0.5 * \h,-1.6) {} ;
    \coordinate (dx2\p) at (1.5 * \h,-1) {} ;
    \coordinate (h12\p) at (1.7 * \h,-1) {} ;
    \coordinate (h22\p) at (1.7 * \h,-1.1) {} ;
    \coordinate (h2b2\p) at (1.7 * \h,-1.12) {} ;
    \coordinate (h2t2\p) at (1.7 * \h,-1.5) {} ;
    \coordinate (h32\p) at (1.9 * \h,-1) {} ;
    \coordinate (y2\p) at (2.5 * \h,-1) {} ;
    \coordinate (py2\p) at (2.5 * \h,-2.8) {} ;
    \coordinate (dy2\p) at (3.5 * \h,-2) {} ;
    \coordinate (e2\p) at (3.9 * \h,-2) {} ;
    \end{scope}

    \end{scope}

    \begin{scope}[very thick]
    \foreach \i in {2}{
      \draw (b\i\p) -- (x\i\p) -- (px\i\p) -- (dx\i\p) -- (h1\i\p) -- (h2\i\p) -- (h3\i\p) -- (y\i\p) -- (py\i\p) -- (dy\i\p) -- (e\i\p) ;
    }
    \draw (e1\p) -- (b2\p) ;
    \end{scope}
    }
    \foreach \p/\i/\j/\h/\k/\s in {2/25/6.8/\f/-2/-4.4}{

      \begin{scope}[xshift=\i cm, yshift= \j cm]
      
    \coordinate (b1\p) at (-0.5 * \h,0) {} ;
    \coordinate (x1\p) at (0.5 * \h,0) {} ;
    \coordinate (px1\p) at (0.5 * \h,-1.3) {} ;
    \coordinate (dx1\p) at (1.5 * \h,-1) {} ;
    \coordinate (h11\p) at (2 * \h,-1) {} ;
    \coordinate (h21\p) at (2 * \h,-1.1) {} ;
    \coordinate (h2b1\p) at (2 * \h,-1.12) {} ;
    \coordinate (h2t1\p) at (2 * \h,-1.5) {} ;
    \coordinate (h31\p) at (2.3 * \h,-1) {} ;
    \coordinate (y1\p) at (3.5 * \h,-1) {} ;
    \coordinate (py1\p) at (3.5 * \h,-2.45) {} ;
    \coordinate (dy1\p) at (4.5 * \h,-2) {} ;
    \coordinate (e1\p) at (4.9 * \h,-2) {} ;

    \begin{scope}[xshift=\s cm,yshift=\k cm]
    \coordinate (b2\p) at (-0.5 * \h,0) {} ;
    \coordinate (x2\p) at (0.5 * \h,0) {} ;
    \coordinate (px2\p) at (0.5 * \h,-1.6) {} ;
    \coordinate (dx2\p) at (1.5 * \h,-1) {} ;
    \coordinate (h12\p) at (1.7 * \h,-1) {} ;
    \coordinate (h22\p) at (1.7 * \h,-1.1) {} ;
    \coordinate (h2b2\p) at (1.7 * \h,-1.12) {} ;
    \coordinate (h2t2\p) at (1.7 * \h,-1.5) {} ;
    \coordinate (h32\p) at (1.9 * \h,-1) {} ;
    \coordinate (y2\p) at (2.5 * \h,-1) {} ;
    \coordinate (py2\p) at (2.5 * \h,-2.8) {} ;
    \coordinate (dy2\p) at (3.5 * \h,-2) {} ;
    \coordinate (e2\p) at (3.9 * \h,-2) {} ;
    \end{scope}

    \end{scope}

    \begin{scope}[very thick]
    \foreach \i in {1,2}{
      \draw (b\i\p) -- (x\i\p) -- (px\i\p) -- (dx\i\p) -- (h1\i\p) -- (h2\i\p) -- (h3\i\p) -- (y\i\p) -- (py\i\p) -- (dy\i\p) -- (e\i\p) ;
    }
    \draw (e1\p) -- (b2\p) ;
    \end{scope}
    }

    \node at (2.6,-4.2) {$T_i$} ;
    \node at (23,2.3) {$T_{i+1}$} ;

    \node at (14,-5) {towards $T_{i-1}$} ;

    \node at (2.3,-2.6) {\tiny{$d_{x,\overline x}$}} ;
    \draw[very thick] (b21)--++(-2.3,0)--++(-0.58,-0.2)--++(0,0.2)--++(-2,0) ; 
    \coordinate (dxx) at (2.1,-2.25) {} ;
    \node[vps] at (dxx) {} ;

    \draw[dashed] (x12) -- (dxx) -- (y12) ;
    
    \begin{scope}[xshift=5.6 cm, yshift=-2 cm]
    \node at (0,-0.5) {$v_y$} ;
    \node at (2,-1.5) {$v_{\overline y}$} ;
    \node at (0.5,-2.1) {$d_y$} ;
    \node at (2.5,-3.3) {$d_{\overline y}$} ;
    \end{scope}

    \begin{scope}[xshift=25 cm, yshift=6.8 cm]
    \node at (0.5 * \f + 0.5,-0.5) {$v_{\overline x}$} ;
    \node at (0.5 * \f,-1.8) {$d_{\overline x}$} ;
    \node at (3.5 * \f + 0.5,-1.5) {$v_x$} ;
    \node at (3.5 * \f,-2.95) {$d_x$} ;

    \begin{scope}[xshift=-4.5 cm, yshift=-2 cm]
    \node at (0.5 * \f + 0.5,-0.5) {$v_{\overline y}$} ;
    \node at (2.5 * \f + 0.5,-1.5) {$v_y$} ;
    \node at (0.5 * \f,-2.1) {$d_{\overline y}$} ;
    \node at (2.5 * \f,-3.3) {$d_y$} ;
    \end{scope}

    \end{scope}

    \begin{scope}[dashed, very thin, red]
    \draw[extended line=2.3cm] (y21) -- (x21) ;
    \draw[extended line=1.2cm] (e21) -- (y21) ;

    \draw[extended line=11.5cm] (y12) -- (x12) ;
    \draw[extended line=9.7cm] (x22) -- (y12) ;
    \draw[extended line=8cm] (y22) -- (x22) ;
    \draw[extended line=6cm] (e22) -- (y22) ;
    \end{scope}

    \begin{scope}[dashed]
    \draw (x22) -- (px21) ;
    \draw (y22) -- (py21) ;
    \end{scope}

    \foreach \i/\p in {1/2,2/1,2/2}{
      \node[vp] at (x\i\p) {} ;
      \node[vp] at (y\i\p) {} ;
      \node[vp] at (px\i\p) {} ;
      \node[vp] at (py\i\p) {} ;
      \node[vps] at (h2b\i\p) {} ;
    }
  \end{tikzpicture}
  \caption{Downward deletion of the variable $x$ (and propagation of the variable $y$). On chunk $T_{i-1}$, the encoding of variable $x$ has totally disappeared: there is \emph{not} even a $d^{i-1}_{x,\overline x}$.}
  \label{fig:downwardDeletion}
\end{figure}
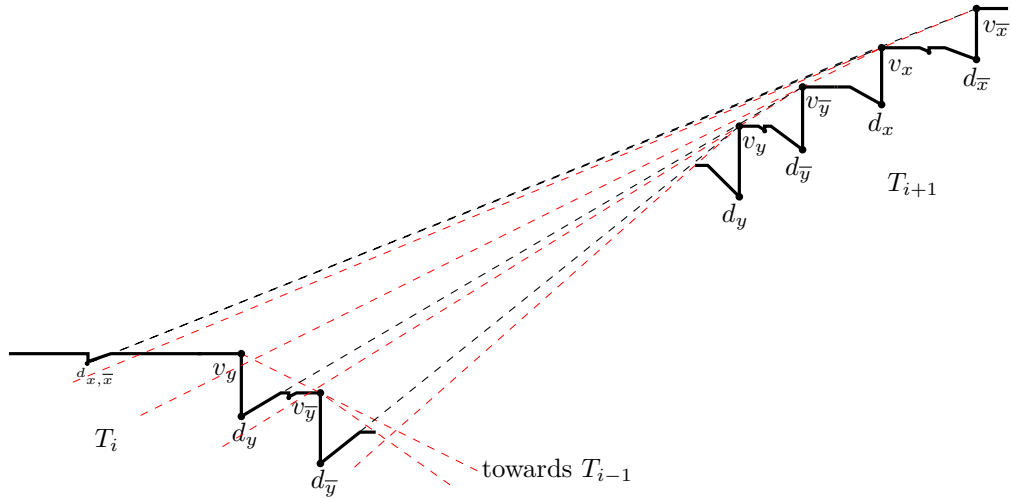

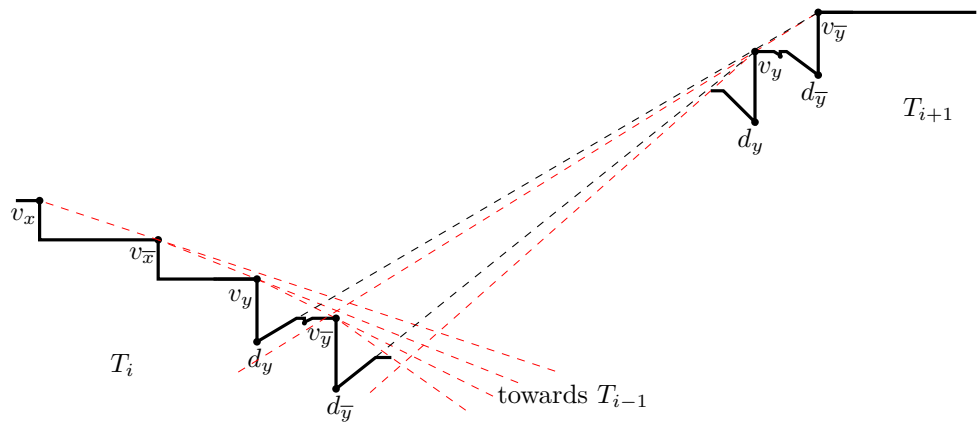
\begin{figure}[h!]
  \centering
  \begin{tikzpicture}
    [scale = 0.52,
      extended line/.style={shorten <= - #1},
      extended line/.default=5cm]

    \def\f{-0.8}

    \foreach \p/\i/\j/\h/\k/\s in {1/0/0/1/-2/5.5,2/25/6.8/\f/-2/-4.4}{

      \begin{scope}[xshift=\i cm, yshift= \j cm]
      
    \coordinate (b1\p) at (-0.5 * \h,0) {} ;
    \coordinate (x1\p) at (0.5 * \h,0) {} ;
    \coordinate (px1\p) at (0.5 * \h,-1.3) {} ;
    \coordinate (dx1\p) at (1.5 * \h,-1) {} ;
    \coordinate (h11\p) at (2 * \h,-1) {} ;
    \coordinate (h21\p) at (2 * \h,-1.1) {} ;
    \coordinate (h2b1\p) at (2 * \h,-1.12) {} ;
    \coordinate (h2t1\p) at (2 * \h,-1.5) {} ;
    \coordinate (h31\p) at (2.3 * \h,-1) {} ;
    \coordinate (y1\p) at (3.5 * \h,-1) {} ;
    \coordinate (py1\p) at (3.5 * \h,-2.45) {} ;
    \coordinate (dy1\p) at (4.5 * \h,-2) {} ;
    \coordinate (e1\p) at (4.9 * \h,-2) {} ;

    \begin{scope}[xshift=\s cm,yshift=\k cm]
    \coordinate (b2\p) at (-0.5 * \h,0) {} ;
    \coordinate (x2\p) at (0.5 * \h,0) {} ;
    \coordinate (px2\p) at (0.5 * \h,-1.6) {} ;
    \coordinate (dx2\p) at (1.5 * \h,-1) {} ;
    \coordinate (h12\p) at (1.7 * \h,-1) {} ;
    \coordinate (h22\p) at (1.7 * \h,-1.1) {} ;
    \coordinate (h2b2\p) at (1.7 * \h,-1.12) {} ;
    \coordinate (h2t2\p) at (1.7 * \h,-1.5) {} ;
    \coordinate (h32\p) at (1.9 * \h,-1) {} ;
    \coordinate (y2\p) at (2.5 * \h,-1) {} ;
    \coordinate (py2\p) at (2.5 * \h,-2.8) {} ;
    \coordinate (dy2\p) at (3.5 * \h,-2) {} ;
    \coordinate (e2\p) at (3.9 * \h,-2) {} ;
    \end{scope}

    \end{scope}

    \begin{scope}[very thick]
    \foreach \i in {2}{
      \draw (b\i\p) -- (x\i\p) -- (px\i\p) -- (dx\i\p) -- (h1\i\p) -- (h2\i\p) -- (h3\i\p) -- (y\i\p) -- (py\i\p) -- (dy\i\p) -- (e\i\p) ;
    }
    \draw (e1\p) -- (b2\p) ;
    \end{scope}
    }

    \node at (2.6,-4.2) {$T_i$} ;
    \node at (23,2.3) {$T_{i+1}$} ;

    \node at (14,-5) {towards $T_{i-1}$} ;

    \node at (0.1,-0.4) {$v_x$} ;
    \node at (3.1,-1.4) {$v_{\overline x}$} ;

    \node[vp] at (x11) {} ;
    \node[vp] at (y11) {} ;

    \draw[very thick] (x11)--++(-0.6,0) ;
    \draw[very thick] (x11)--++(0,-1) -- (y11)--++(0,-1) -- (x21) ;
    \draw[very thick] (x22)--++(4,0) ;

    
    \begin{scope}[xshift=5.6 cm, yshift=-2 cm]
    \node at (0,-0.5) {$v_y$} ;
    \node at (2,-1.5) {$v_{\overline y}$} ;
    \node at (0.5,-2.1) {$d_y$} ;
    \node at (2.5,-3.3) {$d_{\overline y}$} ;
    \end{scope}

    \begin{scope}[xshift=25 cm, yshift=6.8 cm]

    \begin{scope}[xshift=-4.5 cm, yshift=-2 cm]
    \node at (0.5 * \f + 0.5,-0.5) {$v_{\overline y}$} ;
    \node at (2.5 * \f + 0.5,-1.5) {$v_y$} ;
    \node at (0.5 * \f,-2.1) {$d_{\overline y}$} ;
    \node at (2.5 * \f,-3.3) {$d_y$} ;
    \end{scope}

    \end{scope}

    \begin{scope}[dashed, very thin, red]
    \draw[extended line=5.5cm] (y11) -- (x11) ;
    \draw[extended line=3.7cm] (x21) -- (y11) ;
    \draw[extended line=2.3cm] (y21) -- (x21) ;
    \draw[extended line=1.2cm] (e21) -- (y21) ;

    \draw[extended line=8cm] (y22) -- (x22) ;
    \draw[extended line=6cm] (e22) -- (y22) ;
    \end{scope}

    \begin{scope}[dashed]
    \draw (x22) -- (px21) ;
    \draw (y22) -- (py21) ;
    \end{scope}

    \foreach \i/\p in {2/1,2/2}{
      \node[vp] at (x\i\p) {} ;
      \node[vp] at (y\i\p) {} ;
      \node[vp] at (px\i\p) {} ;
      \node[vp] at (py\i\p) {} ;
      \node[vps] at (h2b\i\p) {} ;
    }
  \end{tikzpicture}
  \caption{Upward deletion of the variable $x$ (and propagation of the variable $y$). On chunk $T_{i-1}$ is the usual encoding of variable $x$ with three right triangular pockets.}
  \label{fig:upwardDeletion}
\end{figure}

On all the chunks below the downward deletion of a variable $x$, there is no encoding of variable $x$.  
And, on all the chunks above the upward deletion of a variable $x$, there is no encoding of variable $x$.
The gadgets are represented in Figure~\ref{fig:downwardDeletion} and Figure~\ref{fig:upwardDeletion}, respectively.

This ends the list of gadgets.
The gadgets are assembled as in the reduction of King and Krohn.
From the initial chunk $T_0$ and going up (resp.~going down), one realizes step by step (chunk by chunk) the elementary operations to check the clauses of $\mathcal C^+$ (resp.~$\mathcal C^-$) in the order $C^+_1, C^+_2, \ldots C^+_s$ (resp.~$C^-_1, C^-_2, \ldots C^-_{m-s}$) including propagation, inversion of literals, upward clause checking (resp.~downward clause checking), and upward variable deletion (resp.~downward variable deletion).
Each chunk has $O(n)$ vertices.
Each clause takes $O(1)$ chunks to be checked.
So the total number of chunks is $O(m)=O(n)$ and the total number of vertices is $O(n^2)$.

The total budget is fixed as one per right triangular pocket, two per general triangular pocket, one per variable encoding including the slightly different one at inverters and the one just before an upward deletion (see encoding of variable $x$ on chunk $T_i$ in Figure~\ref{fig:upwardDeletion}), and one extra per inverter.
Note that the lone $d^{\bullet}_{x,\overline x}$ at downward clause gadget and downward deletion do not count as variable gadget and they do not increase the budget.
To give an unambiguous definition of the number of variable encodings, we count the number of pairs $i,x$ such that the vertices $v^i_x$ and $v^i_{\overline x}$ exist. 

We explained why the guards inside the triangular pockets can be placed (and the budget reduced).
At this point the correctness of the reduction is similar to the one by King and Krohn.
Therefore we just sketch it.
The $d$-vertices force to place at least one guard in each variable encoding.
We argued that this will be sufficient to see all the right triangular and rectangular pockets if and only if the variable assignments are consistent between two consecutive chunks (by completing with guards $g^i_{\ell}$ at each inverter where $\ell$ is the literal chosen to be true).
Finally, the terrain is entirely seen whenever the $m$ general triangular pockets corresponding to the $m$ clauses are all guarded, which happens if and only if the \emph{truth assignment} chosen on chunk $T_0$ satisfies all the clauses.  

This shows that \otg and \textsc{Dominating Set} on the visibility graph of rectilinear terrains are NP-hard.
Recall that the continuous variant of \otg is equivalent to its discrete counterpart.
The membership in NP of all those variants is therefore trivial.
What is left to prove is that \textsc{Dominating Set} on the visibility graph of \emph{strictly} rectilinear terrains is NP-hard.
Our reduction almost directly extends to this variant.
The only issue is with the general triangular pocket gadget.
Indeed, when the two guards are placed inside the pocket, all the internal vertices are guarded.
In \otg, one still needed to see the interior of the tiny top horizontal edge.
But this is no longer required in \textsc{Dominating Set}.
We observe that the general triangular pocket is only used in the downward clause gadget.
We explain how we can make the downward clause gadget without the general triangular pocket.
From the gadget depicted on Figure~\ref{fig:downward-clause}, we make the following modifications.
The three literals of the clause are now at positions $2$, $4$, and $5$ on chunk $T_i$.
The \emph{third} literal, that is, the one of the middle variable which does not satisfy the clause has its $v$-vertex slightly lowered in such a way that it does not see anything meaningful on chunk $T_{i-1}$.
On chunk $T_{i-1}$, the right triangular pocket with bottom $d^{i-1}_{y,\overline y}$ is simply removed, and the triangular pocket with bottom $w_C$ is replaced by a right triangular pocket which sees among the $v$-vertices $v^i_{\ell_1}, v^i_{\ell_2}, v^i_{\ell_3}$ and nothing else, for $C=\ell_1 \lor \ell_2 \lor \ell_3$.

What we lose with this new construction is the vertex $d^{i-1}_{y,\overline y}$ which forced to take one $v$-vertex between $v^i_y, v^i_{\overline y}$.
We can now place no guard at those vertices, provided that we place two guards at $v^{i+1}_y$ and $v^{i+1}_{\overline y}$.
However, this can only help if there is also a downward clause gadget between chunks $T^{i+1}$ and $T_i$.
Therefore, we just have to observe the rule of not putting two downward clause gadgets in a row (for instance by separating them with some simple propagation).

\section{Improved ETH-Hardness for (Orthogonal) Terrain Guarding}\label{sec:eth}

\begin{figure}[h!]
  \centering
  \begin{tikzpicture}
    \foreach \i/\j/\k/\h in {0/0/$\xi$/x,-3/0/$a_1$/a1,3/0/$a_2$/a2,0/3/$b_1$/b1,0/-3/$b_2$/b2,-1/1/$\gamma$/c,1/1/$\beta$/b,1/-1/$\alpha$/a,-1/-1/$\delta$/d}{
      \node[draw,circle] (\h) at (\i,\j) {\k} ;
    }
    \foreach \i/\j/\h in {-1.5/1.5/c1,1.5/1.5/c2,1.5/-1.5/c3,-1.5/-1.5/c4,1/0/z1,0/1/z2,-1/0/z3,0/-1/z4,0.6/0/z5,-0.6/0/z6,-2/-0.5/y1,-2/0.5/y2,-0.5/2/y3,0.5/2/y4,2/0.5/y5,2/-0.5/y6,0.5/-2/y7,-0.5/-2/y8}{
      \node[draw,circle,inner sep=-0.05cm] (\h) at (\i,\j) {} ;
    }
    \draw (a1) -- (c1) -- (b1) -- (c2) -- (a2) -- (c3) -- (b2) -- (c4) -- (a1) ;
    \draw (c1) -- (c) ;
    \draw (c2) -- (b) ;
    \draw (c3) -- (a) ;
    \draw (c4) -- (d) ;
    \draw (z1) -- (b) -- (z2) -- (c) -- (z3) -- (d) -- (z4) -- (a) -- (z1) ;
    \draw (a) -- (z5) -- (x) -- (z6) -- (c) ;
    \draw (z6) -- (d) ;
    \draw (z5) -- (b) ;
    \draw (c) -- (y2) -- (a1) -- (y1) -- (d) ;
    \draw (b) -- (y4) -- (b1) -- (y3) -- (c) ;
    \draw (a) -- (y6) -- (a2) -- (y5) -- (b) ;
    \draw (d) -- (y8) -- (b2) -- (y7) -- (a) ;

    \node (l) at (-4,0) {} ;
    \node (r) at (4,0) {} ;

    \begin{scope}[very thick,blue]
      \draw (l) -- (a1) ;
      \path (a1) edge [bend right] (c) ;
      \path (c) edge [bend left=12] (b1) ;
      \path (b1) edge [bend left=12] (b) ;
      \draw (b) -- (x) -- (d) ;
      \path (d) edge [bend right=12] (b2) ;
      \path (b2) edge [bend right=12] (a) ;
      \path (a) edge [bend left] (a2) ;
      \draw (a2) -- (r) ;
    \end{scope}
  \end{tikzpicture}
\caption{The crossover gadget of Lichtenstein for the crossing edges $a_1a_2$ and $b_1b_2$. The large labeled nodes represent variables, and the small unlabeled nodes represent clauses. The clauses ensure that the value of $a_1$ and $a_2$ (resp.~$b_1$ and $b_2$) are the same. The thick blue curved line delimits on one side, the clauses of $\mathcal C^+$, and on the other side, the clauses of $\mathcal C^-$.}
\label{fig:crossover}
\end{figure}
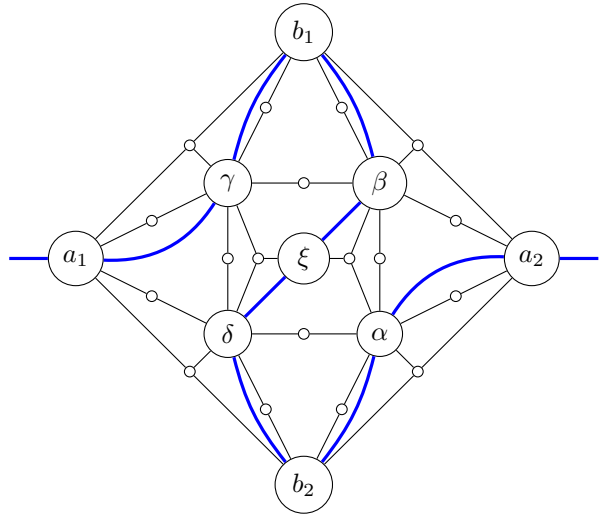

We now explain how to turn the quadratic reductions from \textsc{Planar 3-SAT} into cubic reductions from \textsc{3-SAT} by taking a step back.
This step back is the reduction from \textsc{3-SAT} to \textsc{Planar 3-SAT} by Lichtenstein \cite{Lichtenstein82}, or rather, the instances of \textsc{Planar 3-SAT} it produces.
The idea of Lichtenstein in his classic paper is to replace each intersection of a pair of edges in the incidence graph of the formula by a constant-size planar gadget, called crossover gadget (see Figure~\ref{fig:crossover}).

\begin{figure}[h!]
  \centering
  \begin{tikzpicture}
    \def\n{9}
    \def\s{0.3}
    \foreach \i in {1,...,\n}{
      \draw[opacity=0.2] (\i,1) -- (\i,\n) ;     
      \draw[opacity=0.2] (1,\i) -- (\n,\i) ;
    }

    \foreach \i/\j in {9/1,8/3,7/8,6/4,5/7,4/9,3/2,2/5,1/6}{
      \draw[very thick] (0.5,\i) -- (\j,\i) -- (\j,0.5) ;
    }
    \foreach \i/\j in {1/1,1/2,1/3,1/4,1/5,1/6,1/7,1/8,2/1,2/2,3/1,3/2,3/3,3/4,3/5,3/6,3/7,4/1,4/2,4/3,4/4,4/5,5/1,7/1,7/4,8/1,8/4,8/5,9/1}{
      \draw[thick] (\i+\s,\j) -- (\i,\j+\s) -- (\i - \s,\j) -- (\i,\j - \s) -- (\i+\s,\j) ;
      \path[blue] (\i - \s,\j) edge [bend right=35] (\i,\j+\s) ;
      \path[blue] (\i,\j+\s) edge [bend left=45] (\i,\j) ;
      \path[blue] (\i,\j) edge [bend right=45] (\i,\j - \s) ;
      \path[blue] (\i,\j - \s) edge [bend left=35] (\i+\s,\j) ;
    }

    \foreach \i/\j in {1/8,1/6,1/4,1/2}{
      \path[blue] (\i - \s,\j) edge [bend right=25] (\i - \s,\j - 1) ;
    }
    \foreach \i/\j in {3/7,8/5,4/3}{
      \path[blue] (\i + \s,\j) edge [bend left=25] (\i + \s,\j - 1) ;
    }
    \foreach \i/\j/\k in {1/1/2,2/1/3,3/1/4,4/1/5,5/1/7,7/1/8,8/1/9,1/2/2,2/2/3,3/2/4,1/3/3,3/3/4,1/4/3,3/4/4,4/4/7,7/4/8,1/5/3,3/5/4,4/5/8,1/6/3,1/7/3}{
      \path[blue] (\i + \s,\j) edge [bend left=35] (\k - \s,\j) ;
    }
    \def\t{0.1}
    \def\ins{0.28}
    \foreach \i/\j/\k in {1/1/3,2/4/6,3/7/9}{
    \node (a\i) at (\t,\j) {};
    \node (b\i) at (\t,\k) {};
    \node[draw,rectangle,inner sep=\ins cm,fit=(a\i) (b\i)] {$C_\i$} ;
    }
    \foreach \i/\j/\k/\h in {1/1/2/w,2/3/5/x,3/6/7/y,4/8/9/z}{
    \node (c\i) at (\j,\t) {};
    \node (d\i) at (\k,\t) {};
    \node[draw,rectangle,inner sep=\ins cm,fit=(c\i) (d\i)] (r\h) {$\h$} ;
    }
    \path[blue] (9 + \s,1) edge [bend left=25] (rz.east) ;
    \draw[blue] (rw) -- (rx) -- (ry) -- (rz) ;
    \draw[rounded corners, blue] (1+\s,8) -- (1+\s,9.6) -- (-0.6,9.6) -- (-0.6,0.1) -- (rw.west) ;
  \end{tikzpicture}
\caption{Reduction from \textsc{3-SAT} to \textsc{Planar 3-SAT}, reproduction of Figure 4.3. in Tippenhauer's master thesis \cite{tippenhauer} which follows Lichtenstein's original paper. Notice that some crossover gadgets are used on places without edge intersection, in order to route the blue closed curve (indicating the separation $(\mathcal C^+,\mathcal C^-)$).}
\label{fig:SATtoPlanarSAT}
\end{figure}
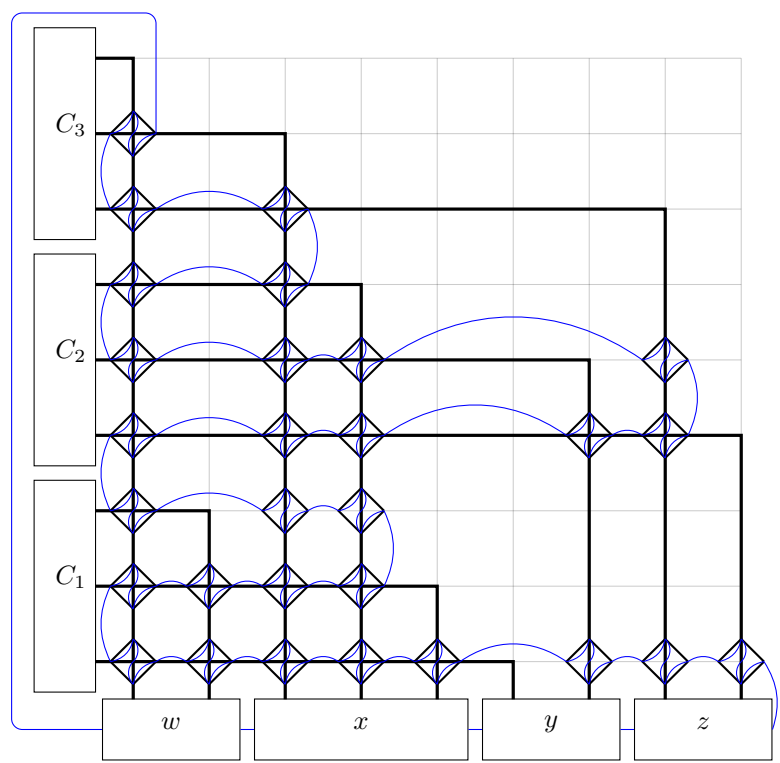

Due to the sparsification of Impagliazzo et al. \cite{Impagliazzo01}, even instances of \textsc{3-SAT} with a linear number of clauses cannot be solved in subexponential time, under the ETH.
Hence, the number of edges in the incidence graph of the formula can be assumed to be linear in the number $N$ of variables.
Thus there are at most a quadratic number $\Theta(N^2)$ of intersections; which implies a replacement of the intersections by a quadratic number of constant-size crossover gadgets.
More concretely, the original $N$ variables (resp.~$\Theta(N)$ clauses) are placed horizontally at the bottom of a $\Theta(N) \times \Theta(N)$ construction grid (resp. vertically at the left of that grid).
Those original variables and clauses are joined in a rectilinear fashion.
Crossover gadgets are placed on a superset of the edge intersections and subset of the grid (see Figure~\ref{fig:SATtoPlanarSAT}).
There is a noose (blue closed curve on the figure) going through all the variables and defining the partition $(\mathcal C^+,\mathcal C^-)$.
Let $\mathcal C^-$ be the part containing the original clauses and $\mathcal C^+$ be the other part.

We wish to reduce the number of chunks that we actually need to check all the clauses.
In the reduction by King and Krohn, each single clause incurs a constant number of chunks: to place the literals at the right position and to check the clause.
However, the only requirement for a clause to be checked is that it operates on consecutive variables.
Therefore, one can check several clauses \emph{in parallel} if they happen to be on disjoint and consecutive variables.
\emph{Checking a set of variable-disjoint clauses in parallel} means that we put the simple propagation/literal inverters/clause gadgets necessary to check a clause, on a constant number of chunks.
In particular, between chunks, say, $T_i$ and $T_{i+1}$, we may have multiple clause checker gadgets.  

A first observation is that the $\Theta(N^2)$ clauses of the crossover gadgets can be checked in parallel with only $O(1)$ chunks.
Indeed, the constant number of clauses within each crossover gadget operates on pairwise-disjoint sets of variables.
They are also consecutive within each gadget with the variable ordering $a_1, \gamma, b_1, \beta, \xi, \delta, b_2, \alpha, a_2$.
We deal first with the remaining clauses of $\mathcal C^+$. 
At this point, there are still potentially $\Theta(N^2)$ equality constraints in $\mathcal C^+$.
In Figure~\ref{fig:SATtoPlanarSAT}, the equality constraints are materialized by thick black edges going from one crossover to another.  
We say that an equality constraint is \emph{vertical} if the corresponding edge contains a vertical section, and that it is \emph{horizontal} otherwise.
Hence a horizontal equality constraint is actually represented by a horizontal segment (without bend).
The \emph{column} of a vertical equality constraint is the column of its (unique) vertical section.

We first check in parallel all the vertical equality constraints of the first column (there are four in Figure~\ref{fig:SATtoPlanarSAT}).
We can then check in parallel all the horizontal equality constraints whose segment ends to the left of the second column (there is just one, on the figure).
Now, the vertical equality constraints of the second column can be checked in parallel (one, in the figure).
We then check at once all the horizontal equality constraints whose segment ends to the left of the third column (three, in the figure), and so on. 
We therefore only need $\Theta(N)$ chunks for $\mathcal C^+$.

For $\mathcal C^-$, we do the same starting from the last column and going down column by column.
After $\Theta(N)$ chunks, we are left with the original variables and clauses which are only $O(N)$.
Thus we finish with $O(N)$ additional chunks.
A chunk contains $O(N^2)$ variable encodings, hence $O(N^2)$ vertices.
So the total number of vertices of a terrain produced from a \textsc{3-SAT} formula on $N$ variables is $O(N^3)$.
This implies that there is no algorithm running in time $2^{o(n^{1/3})}$ for \textsc{(Orthogonal) Terrain Guarding} on terrains with $n$ vertices, unless the ETH fails.

\section{Perspectives}
We have proved that \otg is NP-complete, as well as its variants.
We showed how to get improved ETH-based lower bounds for \tg and \otg, by designing a cubic reduction from \textsc{3-SAT} out of the quadratic reduction from \textsc{Planar 3-SAT}.
This establishes that there is no $2^{o(n^{1/3})}$-time algorithm for those problems, unless the ETH fails.

Besides closing the gap between this lower bound and the existing $2^{O(\sqrt n \log n)}$-algorithm, the principal remaining open questions concern the parameterized complexity of terrain guarding.
\begin{itemize}
\item(1) Is \tg FPT parameterized by the number of guards?
\item(2) Is \otg FPT parameterized by the number of guards?
\end{itemize}
A negative answer to the second question would come as a real surprise in light of the $k^{O(k)}n^{O(1)}$-time algorithm solving \textsc{Dominating Set} on the visibility graph of strictly orthogonal terrains.


\end{document}